\documentclass[useAMS,usenatbib]{mn2e}
\usepackage{graphicx}
\usepackage{aas_macros}
\usepackage{multirow}

\title[Nebulae around Wolf-Rayet Stars]{The spectroscopic properties of the nebulae around seven Galactic and LMC Wolf-Rayet stars } 

\author[D.J. Stock et al.]{D.J. Stock$^{1,2}$\thanks{E-mail: dstock4@uwo.ca (DS)}, M.J. Barlow$^{1}$ and R. Wesson$^1$\\ 
$^{1}$ Department of Physics and Astronomy, University College London, Gower Street, London, WC1E 6BT, UK\\
$^{2}$ Department of Physics and Astronomy, The University of Western Ontario, London ON, N6A 3K7, Canada\\
}

\begin{document}

\date{}

\pagerange{\pageref{firstpage}--\pageref{lastpage}} \pubyear{2011}

\maketitle

\label{firstpage}

\begin{abstract}

We have obtained spectroscopic observations of the nebulae around
seven Wolf-Rayet (WR) stars, four in the Milky Way (WR 8, 16, 18
and 40) and three in the LMC (BAT99-2, -11 and -38). They were
observed using the ESO NTT with the EFOSC-2 instrument, except for
one nebula, NGC 3199, which was observed using the UVES echelle
spectrometer on the VLT. The aims of these observations were to
(a) quantify the degree of chemical enrichment in WR nebulae which
had previously been suggested to have abundances reflecting
nucleosynthetic processing; and (b) to attempt to detect the
far-red lines of neutral carbon (e.g. [C~{\sc i}] 9850~\AA ) in
the nebulae around WC stars. Nebular densities, temperatures and
elemental abundances were derived using standard emission line
diagnostics. For those nebulae for which the usual temperature
diagnostic lines were not detected, their measured [N~{\sc ii}]
and [O~{\sc ii}] line intensities were used to derive reliable
N/O ratios. Our high spectral resolution UVES dataset for NGC 3199
allowed the determination of line broadening temperatures using
lines from several different species. These showed consistent
patterns and fair agreement with the nebular temperatures derived
from diagnostic line ratios.
   Amongst the Galactic WR nebulae, the nebulae around WR 8, 16
and 40 were found to be strongly nitrogen-enriched. NGC~3199 displayed abundances which
were very similar to those of Galactic H~{\sc ii} regions, in
agreement with previous analyses and consistent with the inference
that it is a shell of swept-up interstellar gas. Amongst
the LMC WR nebulae, none displayed
an N/O ratio that significantly exceeded LMC H~{\sc ii} region
values. The far-red [C~{\sc i}] lines were detected
from NGC~3199, with the strength of the 8727\AA\ line indicating 
that at least part of it was produced by C$^+$ recombination rather 
than collisional excitation. The line widths of the [C~{\sc i}] 9824 
and 9850\AA\ lines were however the same as those of collisionally 
excited nebular lines, indicating that those lines originated 
largely from ionized regions within the nebula.

\end{abstract}


\section{Introduction}

Massive stars are known to be prodigious recyclers of the ISM, for example a 60 M$_\odot$ star is thought to end its life as a black hole with a mass of around a few solar masses \citep{2009ARAA..47...63S}, implying that that the star will expel around 90\% of its initial mass into the ISM during its lifetime. This mass loss, either as a smooth wind or a violent episode, can create circumstellar nebulae. We can help to quantify massive star recycling, in terms of its nucleosynthetic effects, by spectroscopically investigating the nebulae they create. 

Spectroscopic observations of WR nebulae have been performed several times in the past, usually focusing on the brightest Galactic examples e.g. \citet{K84, E90, E91,  1992AA...259..629E,  E92-6888, M99}. As summarised by \citet{2010MNRAS.409.1429S}, the nebular abundances found tend to fall into three groups: no enrichments (wind blown bubbles, denoted W type nebulae); helium enrichment (wind blown bubbles with some degree of enrichment, denoted W/E type nebulae) and full blown He and N enrichments which are taken to be evidence for stellar ejecta (ejecta or E type nebulae). 

Massive stars have been mooted as possible sites of significant carbon formation \citep{2000ApJ...541..660H}. Wolf-Rayet stars in their WC phase possess very carbon rich atmospheres and high mass loss rates. Their role in enriching carbon has proven difficult to quantify given the challenge of observing gaseous carbon in nebulae around evolved stars. The collisionally excited lines (CELs) of carbon are elusive, lying in relatively inconvenient parts of the electromagnetic spectrum: e.g. the [C~{\sc ii}] fine structure line in the far IR (158$\mu$m), various lines in the UV (e.g. C~{\sc iii}] 1909\AA)  and [C~{\sc i}] lines in the far red (8700-9900\AA). Observations of the far-IR and UV lines rely upon satellite missions like Herschel and the HST (Hubble Space Telescope). In contrast, the [C~{\sc i}] lines can be accessed from the ground, should they have sufficient intensity.

The far-red carbon lines ([C~{\sc i}] 8727, 9824, 9850\AA) are routinely observed in certain astrophysical contexts, particularly from planetary nebulae and H~{\sc ii} regions. \citet{1995MNRAS.273...47L} observed the lines in PN spectra, detecting all three lines in four nebulae. The lines are also common in the photodissociated regions (PDRs) surrounding H~{\sc ii} regions, e.g. \citet{1982NYASA.395..170M} for NGC 2024, \citet{1992MNRAS.257P...1B} for NGC 2023. However it was subsequently shown that in these PDRs the lines must arise from recombination rather than collisional excitation \citep{1991ApJ...375..630E} which is also the case for some PNe \citep{1973ApL....14..115D} although not for the PNe observed by \citet{1995MNRAS.273...47L}. 

\citet{2010MNRAS.409.1429S} recently surveyed and classified nebulae around WR stars, discovering a new nebula around the WN/WC type star  WR 8 and identifying an LMC WC star whose nebula would be appropriate for an abundance analysis. These nebulae, along with a ``control group" of other WR nebulae were identified and observed with ESO's 3.6m NTT/EFOSC2 (ESO faint object spectrograph and camera) \citep{1984Msngr..38....9B, 2008Msngr.132...18S} in long slit spectroscopy mode with grisms covering regions from 3100-10000\AA\ that included a variety of standard nebular diagnostics. Complementary  service mode observations of NGC 3199 around WR 18 were performed concurrently using ESO's UVES instrument mounted on UT2 of the Very Large Telescope (VLT) \citep{2000SPIE.4008..534D}.

From the data obtained we derive nebular physical conditions using standard nebular density and temperature indicators and then use these to derive elemental abundances from the CELs. We then compare these quantities with previously published physical conditions and abundances for these nebulae and with those predicted by stellar evolution models. For the nebulae which are previously unobserved, we estimate whether there is any stellar ejecta content within the nebular material. 

\section{Target Selection}

The LMC WC star BAT99-11 and the Milky Way WN/WC star WR 8 are carbon rich WR stars with ejecta-type circumstellar nebulae. One of the aims of our observing programme was to attempt to detect the far-red carbon lines ([C~{\sc i}] 8727, 9824, 9850\AA) in the nebulae surrounding both stars. Four other nebulae, each around a nitrogen rich (WN type) WR star, were chosen as a representative sample against which comparisons could be drawn; both with the nebulae with WC central stars and with nebulae analysed by previous authors.  The parameters of each of the chosen targets are summarised in Table~\ref{Targets_Table}.

Four of the targets (WR 16, NGC 3199, RCW58 and BAT99-2) have been the subject of previous spectroscopic analyses, as briefly discussed by \citet{2010MNRAS.409.1429S}. In general helium and nitrogen overabundances were found, however in each case the origin of the ejecta is still subject to debate in that it is possible that the nebular material could have been ejected during the red supergiant (RSG) or luminous blue variable (LBV) phases through which massive stars pass prior to becoming a WR star.  

The final target, the nebulosity thought to be associated with BAT99-38, has not previously been observed spectroscopically. BAT99-38 resides in the LMC superbubble N51D, and therefore has quite complex surroundings \citep{2005ApJ...634L.189C}. \citet{2010MNRAS.409.1429S} identified the arc of nebulosity near BAT99-38 as a possible ejecta nebula despite its presence within a superbubble. 

\begin{table*}
	\centering
	\caption{Target Parameters}	

	\begin{tabular}{lp{3.0cm}p{1.5cm}ccccc}
	\hline
	Nebula & Central Star & Spectral Type$^a$ & v mag.$^a$ & $A_v^a$ & E(B-V)$^a$ & RA(2000) & Dec(2000)\\
	\hline 
	\textit{Galactic} \\
	
	Anon    & WR 8 (HD 62910)   & WN7/WC4 & 10.48 & 2.64    & 0.85 & 07 44 58.2 & -31 54 29.6  \\
	Anon    & WR 16 (HD 86161)  & WN8h    & 8.44  & 2.05    & 0.66 & 09 54 52.9 & -57 43 38.3  \\
	NGC 3199& WR 18 (HD 89358)  & WN4     & 11.11 & 2.92    & 0.94 & 10 17 02.3 & -57 54 46.9  \\	
	RCW 58  & WR 40 (HD 96548)  & WN8h    & 7.85  & 1.56    & 0.50 & 11 06 17.2 & -65 30 35.2  \\
	\\	
	\textit{LMC} \\	
		
	Anon    & BAT99-2 (Brey 2)   & WN2     & 16.22 & $<$ 0.5 &      & 04 49 36.1 & -69 20 54.5  \\
	Anon    & BAT99-11 (HD 32402)& WC4     & 13.95 & $<$ 0.5 &      & 04 57 24.1 & -68 23 57.3  \\  	
	Anon    & BAT99-38 (HD 36402)& WC4+O   & 11.50 & $<$ 0.5 &      & 05 26 03.9 & -67 29 57.0  \\ 	
	\hline	
	\end{tabular}

	\medskip
	\textit{a}: from \citet{H00} (Galactic), \citet{B99} (LMC) 
\label{Targets_Table}
\end{table*}

\subsection{Previous Spectroscopic Observations}

RCW 58 was the subject of several spectroscopic investigations in the years following its identification by \citet{C81} as a potential WR ejecta nebula. \citet{K84} presented the first long slit spectra of the nebulosity, which was found to be highly enriched in helium and nitrogen at an assumed temperature of 7500 K. Subsequently the dynamical structure of the nebula was revealed by \citet{1988MNRAS.234..625S} to be a combination of a quickly expanding shell interspersed with higher density clumps. It was speculated that the higher density material was the product of pre-WR stellar evolution and did not represent ejecta from the WR phase. 

The anonymous circumstellar nebula surrounding WR 16 has been observed by \citet{M99} with surprising results. The nebular spectrum was devoid of several expected lines e.g. [O~{\sc iii}] 4959, 5007\AA ;  [S~{\sc ii}] 6717, 6731\AA.  \citet{M99} suggested that this was due to high densities ($n_e > 10^4$ cm$^{-3}$ ) suppressing these lines. They noted that the [S~{\sc ii}] 6717, 6731\AA\ lines reappear in the outer regions of the nebula, which was assumed to indicate a much lower density of a tenth of the value which they assumed for the inner nebula.

NGC 3199 has been the subject of several spectral analyses, albeit none with comparable resolution to our VLT/UVES dataset. \citet{1992AA...259..629E} found abundances generally in accordance with those of Galactic H~{\sc ii} regions. 

The circumstellar nebula around BAT99-2 was first observed spectroscopically by \citet{1994PASP..106..626G} and subsequently by \citet{2003AA...401L..13N} (the star was referred to in both cases as Brey-2). Both found that the nebula was highly ionized - displaying He~{\sc ii} and [Ar~{\sc iv}] lines which are not normally observed in WR nebulae. BAT99-2 is a WN2 type star \citep{2003MNRAS.338.1025F}, amongst the hottest WR stars, with a surface temperature above 90,000 K \citep{C07}, producing very intense EUV radiation and ionizing its surroundings to a greater degree than the nebulae around other WR stars. No significant abundance enhancements were detected in the nebula by \citet{2003AA...401L..13N}.

\section{Observations}

\subsection{NTT/EFOSC2}

Most of the observations were performed in visitor mode over five days in December 2009 using the EFOSC2 instrument mounted on the 3.6m ESO New Technology Telescope (NTT) at La Silla Observatory, Chile. Two grisms were used in long slit spectroscopy mode providing a wavelength coverage from 3095-5085\AA\ and 6015-10320\AA. The EFOSC2 slit length was 4.1 arcmin which, due to the fact that almost all of our targets had smaller angular sizes than this, enabled us to use the off-target portions of the slit for sky-subtraction. In all science observations the slit width was set to 1 arcsecond, while for standard stars the widest slit (5 arcseconds) was used. 

Grisms 14 (hereafter Blue) and 16 (hereafter Red) respectively have resolutions (FWHM) of 8\AA\ and 14\AA\ respectively. This allowed us to use 2*2 binning throughout our observations without losing any information, as the line profiles were still adequately sampled in this binning mode. This translated to 1030 pixels in both the dispersion and spatial axes.

A log of the nebular observations is presented in Table~\ref{NTTObsJournal}. The slit positions used for each nebula are shown in Figures~\ref{B2_slit}--\ref{WR40_slit}. A summary of the total integration times in both blue and red grisms is shown in Table~\ref{NTTObstots}.

Multiple spectroscopic standard stars were observed each night, notably LTT 1788, HR718, and LTT 3864. Observing templates for these stars existed in the EFOSC2 database and were used each night so as to acquire the appropriate signal to noise level for each object to allow later flux calibration.

\begin{table*}
	\centering
	\caption{Journal of Nebular Observations}	
	\begin{tabular}{llcclcp{1.5cm}p{1.25cm} }
		\hline
		Nebula & Central Star & Date & Instrument & Grism & $\Delta\lambda$ (\AA) & Exposure Time (s) & No. Exposures  \\
		\hline
		Anon    & WR 8     & 2009 Dec. 19/20 & EFOSC2 & 14 & 3095-5085  & 1800 & 2 \\ 
		Anon    & WR 8     & 2009 Dec. 19/20 & EFOSC2 & 16 & 6015-10320 & 1800 & 4 \\
		Anon    & WR 8     & 2009 Dec. 20/21 & EFOSC2 & 14 & 3095-5085  & 1200 & 4 \\
		Anon    & WR 8     & 2009 Dec. 20/21 & EFOSC2 & 16 & 6015-10320 & 1500 & 6  \\		
		Anon    & WR 16    & 2009 Dec. 21/22 & EFOSC2 & 14 & 3095-5085  & 1800 & 3  \\
		Anon    & WR 16    & 2009 Dec. 21/22 & EFOSC2 & 16 & 6015-10320 & 1800 & 3  \\
		NGC 3199 & WR 18    & 2009 Dec. 28/29 & UVES  & DIC1 & 3030-3880 \& 4760-6840  & 1490 & 4\\
		NGC 3199 & WR 18    & 2009 Dec. 28/29 & UVES  & DIC2 & 3730-4990 \& 6600-10600 & 1490 & 2\\
 		NGC 3199 & WR 18    & 2010 Jan. 1/2   & UVES  & DIC2 & 3730-4990 \& 6600-10600 & 1490 & 2\\
		RCW 58  & WR 40    & 2009 Dec. 18/19 & EFOSC2 & 14 & 3095-5085  & 600  & 4  \\
		RCW 58  & WR 40    & 2009 Dec. 18/19 & EFOSC2 & 16 & 6015-10320 & 600  & 5  \\
		RCW 58  & WR 40    & 2009 Dec. 19/20 & EFOSC2 & 14 & 3095-5085  & 600  & 5  \\
		RCW 58  & WR 40    & 2009 Dec. 19/20 & EFOSC2 & 16 & 6015-10320 & 600  & 5  \\
		Anon    & BAT99-2  & 2009 Dec. 22/23 & EFOSC2 & 14 & 3095-5085  & 1200 & 3  \\
		Anon    & BAT99-2  & 2009 Dec. 22/23 & EFOSC2 & 16 & 6015-10320 & 1200 & 3  \\
		Anon    & BAT99-2  & 2009 Dec. 21/22 & EFOSC2 & 14 & 3095-5085  & 1200 & 3  \\
		Anon    & BAT99-2  & 2009 Dec. 21/22 & EFOSC2 & 16 & 6015-10320 & 1200 & 3  \\
		Anon    & BAT99-11 & 2009 Dec. 18/19 & EFOSC2 & 14 & 3095-5085  & 1500 & 4  \\
		Anon    & BAT99-11 & 2009 Dec. 18/19 & EFOSC2 & 16 & 6015-10320 & 1500 & 8  \\
		Anon    & BAT99-11 & 2009 Dec. 22/23 & EFOSC2 & 14 & 3095-5085  & 1800 & 3  \\
		Anon    & BAT99-11 & 2009 Dec. 22/23 & EFOSC2 & 16 & 6015-10320 & 1800 & 6  \\
		Anon    & BAT99-38 & 2009 Dec. 19/20 & EFOSC2 & 14 & 3095-5085  & 900  & 3 \\
		Anon    & BAT99-38 & 2009 Dec. 19/20 & EFOSC2 & 16 & 6015-10320 & 900  & 3  \\
		\hline
	\end{tabular}
\label{NTTObsJournal}	
\end{table*}

\begin{table}
	\centering
	\caption{Total integration times by target}	
	\label{NTTObstots}
	\begin{tabular}{p{1.5cm}p{1.5cm}p{1.5cm}p{1.5cm}}
	\hline	
	Nebula & Central Star & Total Red (hours) & Total Blue (hours)\\
	\hline
	Anon    & WR 8     & 4.5       & 2.33      \\
	Anon    & WR 16    & 1.5       & 1.5       \\
	NGC 3199 & WR 18    & 1.6       & 1.6       \\	
	RCW 58  & WR 40    & 1.5       & 1.66      \\
	Anon    & BAT99-2  & 2         & 2         \\
	Anon    & BAT99-11 & 6.33      & 3.16      \\  	
	Anon    & BAT99-38 & 0.75      & 0.75      \\ 	
	\hline
	\end{tabular}
\end{table}

\subsection{VLT/UVES} \label{UVES_int_sec}

Service mode VLT/UVES observations of NGC 3199 were peformed almost concurrently with the NTT/EFOSC2 observations (see Table~\ref{NTTObsJournal}). These were designed to assist with the main science goal while also potentially detecting heavy element recombination lines. The data cover the wavelength region from 3000-10400\AA\ with only small gaps between 5700-5800\AA\ and 8500-8650\AA\ due to small gaps between CCD chips. The UVES dataset had a resolving power of $\frac{\lambda}{\delta\lambda} = 40000$ with a 0.6 arcsecond slit width and a 15 arcsecond slit length. Its high resolution allowed us to forego the sky subtraction process, as night-sky emission lines appear at the instrumental resolution as opposed to the nebular lines for which we could measure their true widths.

\subsection{Individual Objects}

\subsubsection*{BAT99-2}

\begin{figure}
	\centering
	\includegraphics[width=85mm]{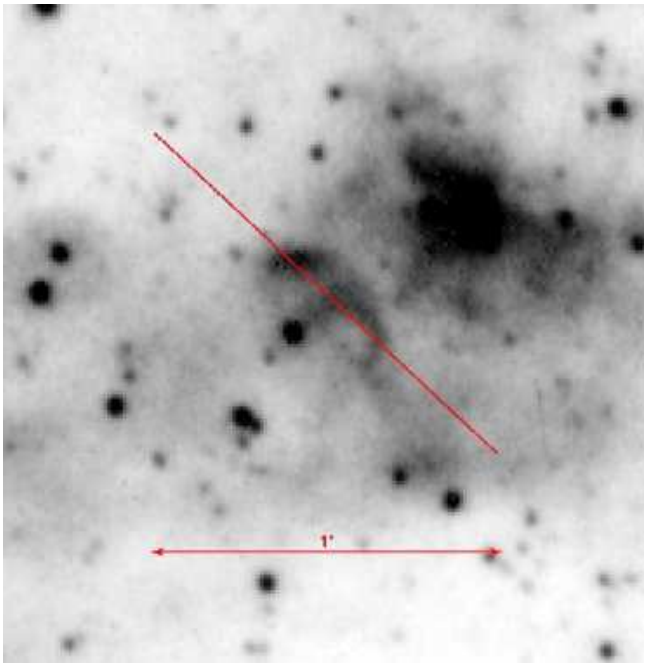}	
	\caption{SHS H$\alpha$ image \citep{P05} showing the location of the slit for the nebula around BAT99-2. Entire slit length (5') is not shown.}
	\label{B2_slit}

\end{figure}

The slit position adopted for the LMC WR nebula around BAT99-2 is shown in Figure~\ref{B2_slit}. This position was adopted as it placed the greatest extent of the arc shaped nebulosity under the EFOSC2 slit. The slit position we adopted roughly matches the position described as ``arc" by \citet{2003AA...401L..13N}. 

\subsubsection*{BAT99-11}

\begin{figure}
	\centering
	\includegraphics[width=85mm]{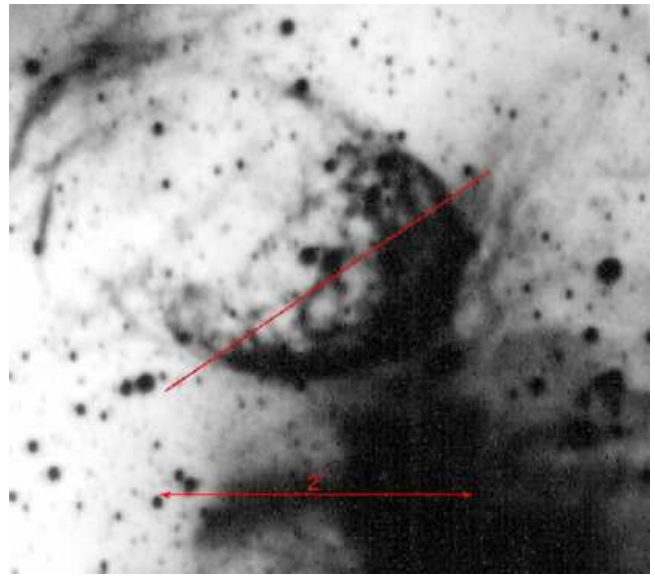}
	\caption{SHS H$\alpha$ image showing the location of the slit for the nebula around BAT99-11. Entire slit length (5') is not shown.}
	\label{B11_slit}
\end{figure}

The slit position adopted for the nebula around BAT99-11 in the LMC is shown in Figure~\ref{B11_slit}. The nebulosity around BAT99-11 is physically much larger than that normally associated with WR ejecta nebulae, so the slit position was chosen to intersect the inner portions of the nebula as this may be more recently expelled material.

\subsubsection*{BAT99-38}

\begin{figure}
	\centering
	\includegraphics[width=85mm]{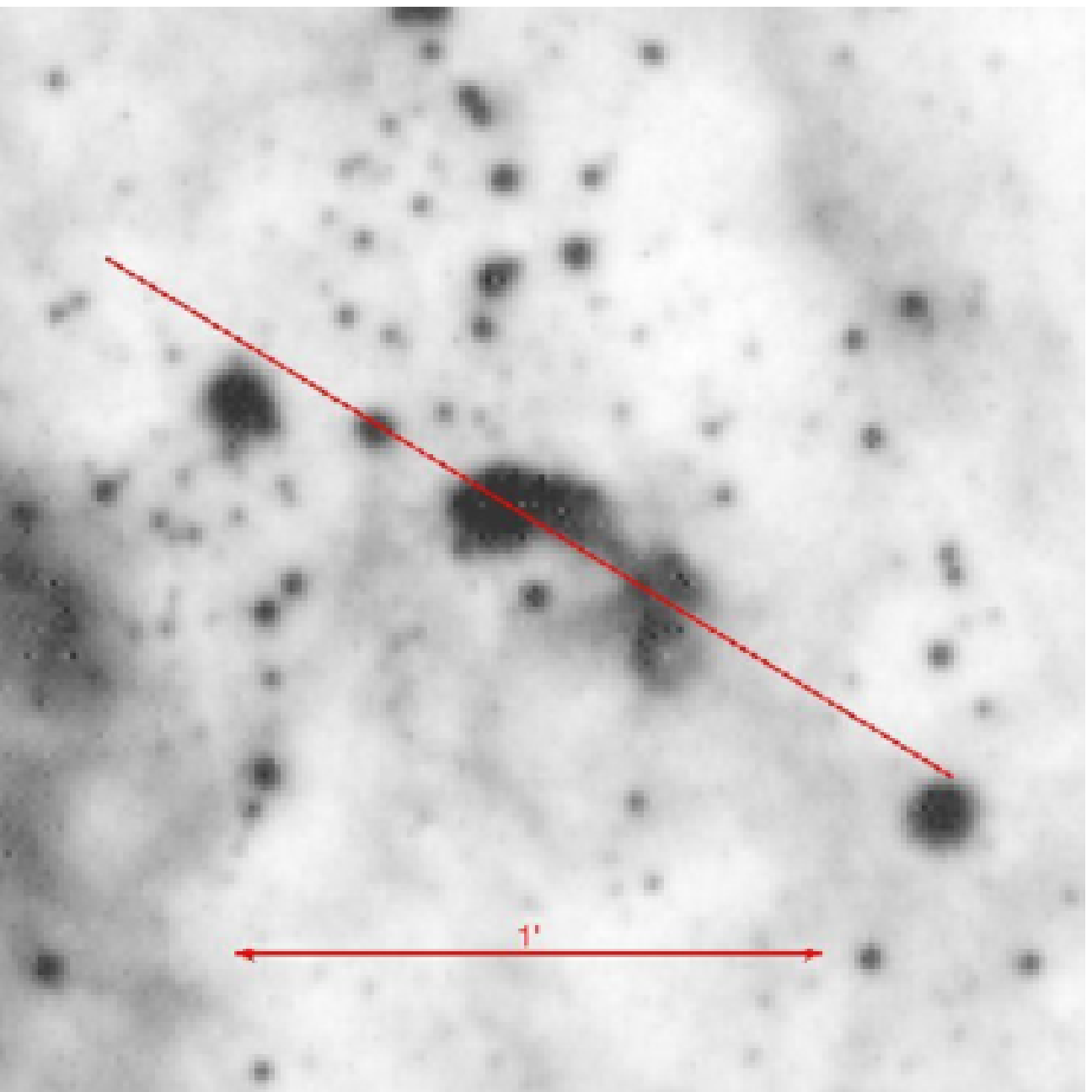}
	\caption{SHS H$\alpha$ image showing the location of the slit for the nebulosity near  BAT99-38. Entire slit length (5') is not shown.}
	\label{B38_slit}
\end{figure}

The slit position adopted for the nebulosity near BAT99-38 is shown in Figure~\ref{B38_slit}. It was discovered later that BAT99-38 was not the star which appears at the focus of the arc, but is embedded in the arc itself. This was due to a small error in the WCS system of the digitised SHS survey plates which were used to identify WR nebulae in the LMC (see \citet{2010MNRAS.409.1429S}). The slit position therefore unexpectedly included BAT99-38 along with the nebulosity. The slit was subseqently moved slightly such that BAT99-38 was no longer included in the slit. The position of the star called into question our preliminary morphological identification of this nebula as a possible ejecta nebula. 

\subsubsection*{WR 8}

\begin{figure}
	\centering
	\includegraphics[width=85mm]{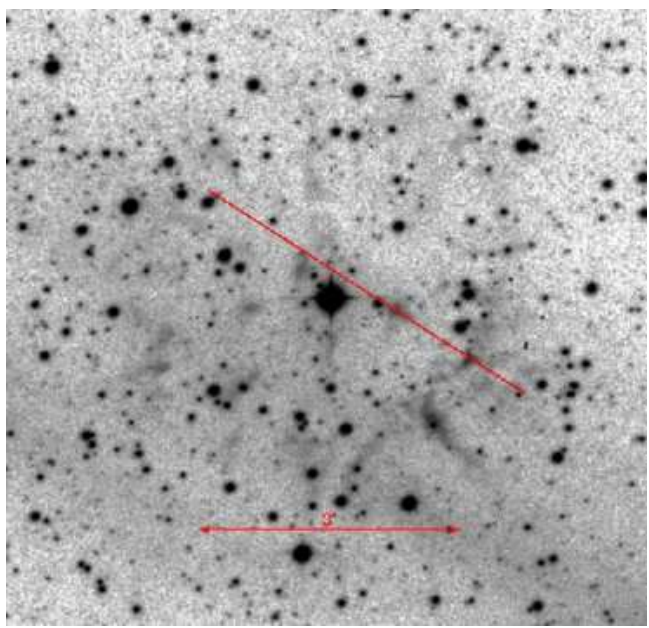}
	\caption{SHS H$\alpha$ image showing the location of the slit for the nebula around WR 8. Entire slit length (5') is not shown.}
	\label{WR8_slit}
\end{figure}

As with BAT99-11, it is unclear whether the knots which appear interior to the outer ring of nebulosity are in the same shell as the visible ring, or are really spatially closer to WR 8. The slit position was chosen to encompass the two brightest H$\alpha$ emitting knots which did not coincide with stars. The slit position adopted is shown in Figure~\ref{WR8_slit}.

\subsubsection*{WR 16}

\begin{figure}
	\centering
	\includegraphics[width=85mm]{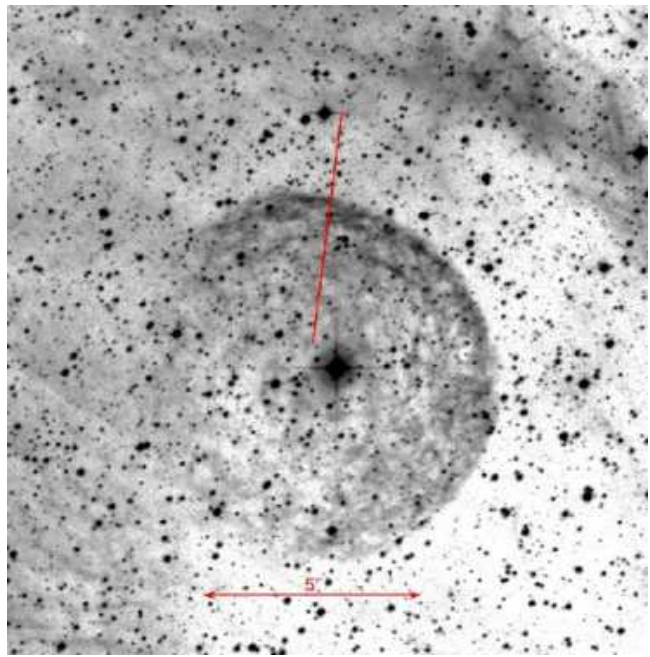}
	\caption{SHS H$\alpha$ image showing the location of the slit for the nebula around WR 16. }
	\label{WR16_slit}
\end{figure}

The adopted slit position for WR 16 is shown in Figure~\ref{WR16_slit}. Due to the large size of this nebula the slit shown in Figure~\ref{WR16_slit} is appoximately the true length of the EFOSC2 slit ($\sim$ 5'). The slit position was chosen such that it encompassed some of the brighter H$\alpha$ knots seen in the SHS imagery, along with enough off-nebula sky to perform a reasonable sky subtraction.

\subsubsection*{NGC 3199 (WR 18)}

\begin{figure}
	\centering
	\includegraphics[width=85mm]{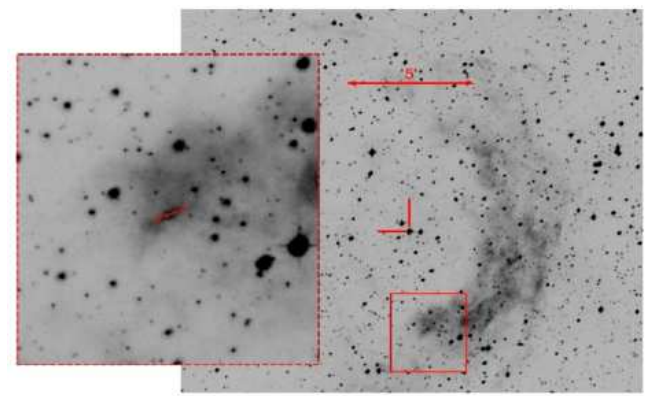}
	\caption{SHS short red image of NGC 3199 (location of WR 18 indicated with dashes), $\sim$ 15 arcsecond VLT/UVES slit is shown in the expanded section.}
	\label{NGC3199_slit}
\end{figure}

The slit position adopted for NGC 3199 is shown in Figure~\ref{NGC3199_slit}. Given the shorter (15") UVES slit length the brightest knot visible in the short red SHS exposure was chosen. For this object the H$\alpha$ SHS image is saturated and so not useful in terms of selecting the brightest parts of the nebulosity. While NGC 3199 is very much brighter in H$\alpha$ than our other targets, it was still important to select the brightest possible section of nebulosity such that the likelihood of detecting weak lines was maximised.    

\subsubsection*{RCW 58 (WR 40)}

\begin{figure}
	\centering
	\includegraphics[width=85mm]{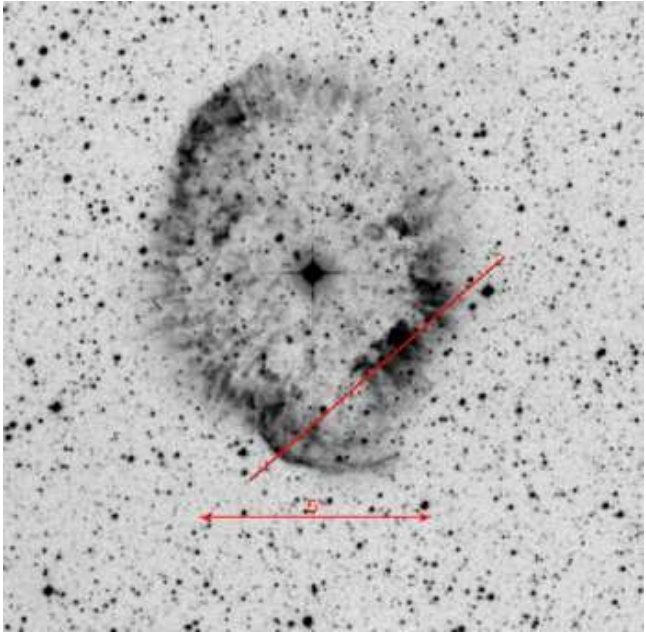}
	\caption{SHS H$\alpha$ image showing the location of the slit for RCW 58 (WR 40).}
	\label{WR40_slit}
\end{figure}

The slit position used for observations of RCW 58 is shown in Figure~\ref{WR40_slit}. This slit position was adopted to maximise the amount of bright nebulosity on the western side of RCW 58 in the slit.

\section{Data Reduction}\label{DR_sec}

\subsection{EFOSC2 Spectra}

\begin{figure}
	\centering
	\includegraphics[width=85mm]{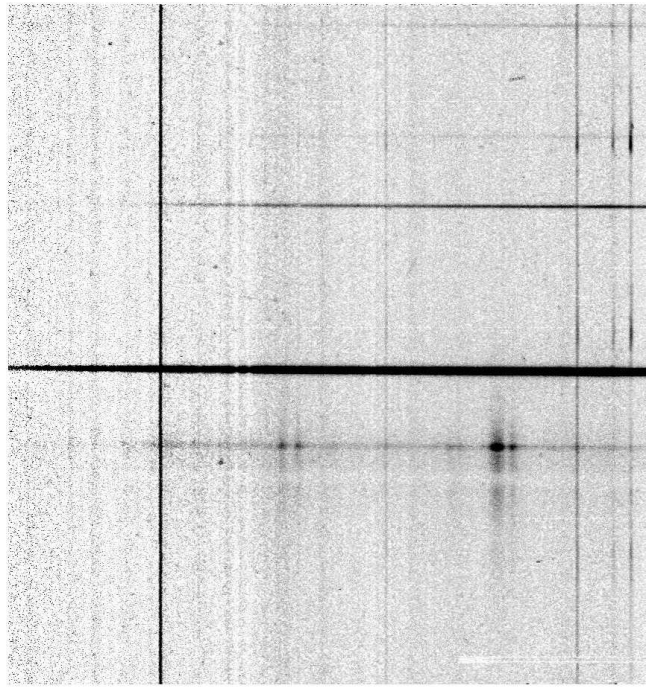}
	\caption{2D Spectrogram of the nebulosity around WR 8 in the 3000-5050\AA\ wavelength region (Grism \#14). The nebular lines at the top are from the nebulosity shown in Figure~\ref{WR8_slit}. The emission line source in the lower portion of this spectrum is the scattered light of the WR central star. }
	\label{WR8_2D}
\end{figure}

 At the beginning and end of each night a series of calibration frames were acquired, including bias, arc and flat field frames. The reduction of the data was performed using ESO-MIDAS \citep{1992ASPC...25..115W} and Gaussian line measurements were performed using the ELF emission line fitting routine written by P.J. Storey within the Dipso Starlink package written by I. Howarth. A typical result of the data reduction process, a 2D spectral image showing nebular lines, is shown in Figure~\ref{WR8_2D}

\begin{figure*}
	\centering
	\includegraphics[width=180mm]{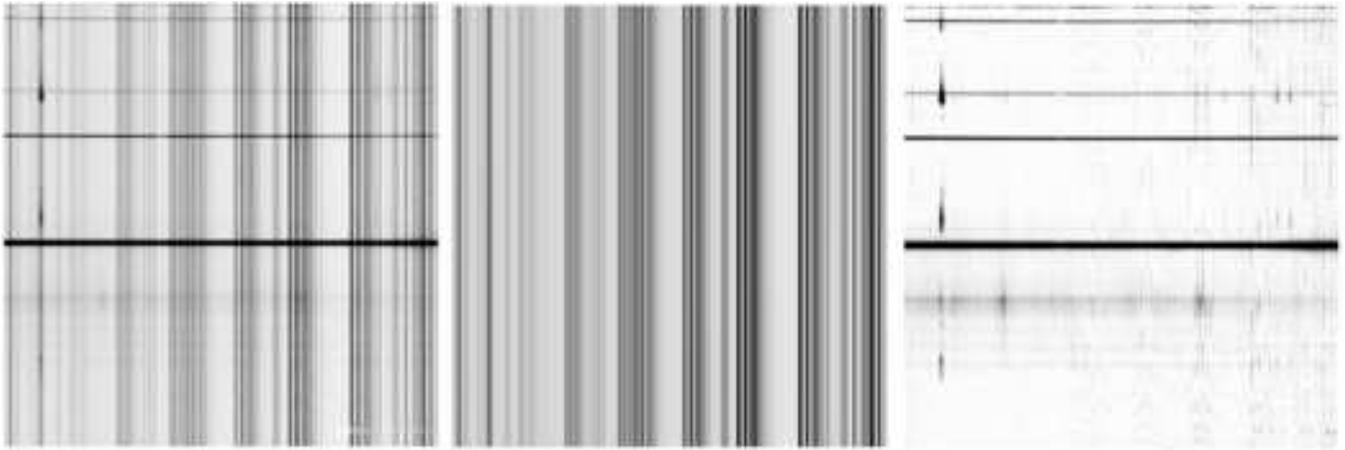}
	\caption{ \textit{left:} 2D spectra of the region around WR 8 for the red spectral region (6000-10000\AA\, Grism \#16); \textit{centre:} ESO-MIDAS ``SKYFIT/LONG" generated 2D sky spectrum for the frame shown; \textit{right:} Sky subtracted spectrogram of the region around WR 8.}
	\label{SS}
\end{figure*}

Sky subtraction was performed using the ESO-MIDAS routine ''SKYFIT/LONG", which, given two sky windows either side of the object spectrum, calculates a 2D frame of the night sky lines by interpolating across between the windows. This can then be subtracted from the data frame to leave just the object spectrum in the resulting frame. The 7000-9000\AA\ region is particularly blighted by night sky lines, as shown in the composite Figure~\ref{SS}, which displays a non-sky subtracted frame, an interpolated sky frame and the sky-subtracted image. Post sky-subtraction, some evidence of fringing appears which we were not able to remove using any combination of internal flats or dome flats. This residual fringing is not thought to be a problem as there were very few lines being measured in the 7000-9000\AA\ spectral region. 

For each 2D co-added sky-subtracted nebular spectrum, every row of pixels which was thought to contain an uncontaminated nebular spectrum was co-added to produce 1D spectra. Subsequently each line was measured using Dipso/ELF.

The reddening was derived using the Balmer series ratios to calculate c(H$\beta$) - the logarithmic extinction at H$\beta$ - using pairs of Balmer lines (one of which was always H$\beta$). The derived reddening values (Table~\ref{Redtab}) were consistent with literature values derived by previous authors in the cases of WR 16 \citep{M99}, WR 40 \citep{K84} and BAT99-2 \citep{2003AA...401L..13N}. They were also checked against the Galactic extinction maps of \citet{1998ApJ...500..525S}. In every case the Galactic WR nebulae had derived E(B-V) values lower than the Schlegel et al maximum value (indicating that they are, in fact, resident in the Galaxy - a useful sanity check). However for the LMC nebulae the Schlegel dust maps indicated E(B-V) values which were too high by at least a factor of five. For LMC nebula we resorted to the \citet{1982AJ.....87.1165B} E(B-V) maps which were found to agree much better with our data. 

In some cases the dereddening process highlighted a lack of a reliable relative flux calibration between the blue and red spectra. For the nebulae around WR 16 and BAT99-11, this manifested itself as very different c(H$\beta$)'s when comparing H$\alpha$ in the red spectra to H$\beta$ in the blue, compared with those derived from just the blue spectra (using H$\delta$ and H$\gamma$). For the nebula around WR 8, the converse situation applied in that a very low c(H$\beta$) was found from the blue Balmer lines and a high c(H$\beta$) was found from H$\alpha$. This may have been caused by changing conditions throughout our allocated nights, making the flux calibrations derived for the beginning and end of each night inappropriate for the middle of each night. This effect was minimised by using the nearest (in time) flux calibration to each observation. The adopted c(H$\beta$) for both blue and red settings, where they did not agree, are shown in Table~\ref{Redtab}. In cases where a separate reddening solution was adopted for the red specta, the calibration scales the red spectra such that H$\alpha$ has the correct flux ratio to the Balmer lines in the blue spectra. This has the consequence of ensuring that the region around H$\alpha$ also posesses the correct relative flux levels (e.g. [N~{\sc ii}], [S~{\sc ii}]). Spectral lines detected far from H$\alpha$ were considered unreliable in these cases.

A futher complication was intoduced by the choice to deviate from the parallactic angle in order to maximise the amount of nebulosity falling under the slit. In cases where the nebulae were large and of fairly constant surface brightness, this did not cause a problem. However, the nebulae with discordant blue/red reddening values were those where the slit passed through several small clumps, e.g. WR 8 (See Figure~\ref{WR8_slit}). In these cases the effects of atmospheric dispersion combined with the small angular scales of the clumps of nebulosity may be the cause of the reddening issues. The offsets from the parallactic angle for each object are listed in Table~\ref{NTTObsJournal}. 

By adopting a different reddening solution for the red spectra we effectively de-redden and flux calibrate H$\alpha$ and the adjacent lines ([N~{\sc ii}], [S~{\sc ii}], He~{\sc i} 6678\AA) but lose flux calibration for lines far from H$\alpha$ ([S~{\sc iii}] 9069, 9531\AA\ for example). For the nebulae which required this technique (those around WR 8, WR 16 and BAT99-11), this would render their [S~{\sc iii}] diagnostic temperatures unreliable.

\begin{table*}
	\centering
	\caption{Derived Reddenings}	
	\label{Redtab}		
	\begin{tabular}{p{1.4cm}p{1.4cm}p{1.0cm}p{1.1cm}p{1.1cm}p{1.15cm}p{1.15cm}p{1.15cm}p{1.5cm}p{2.8cm} }
	\hline	
	Nebula & Central Star & c(H$\beta$) & c(H$\beta$)$_{blue}^{a}$ & c(H$\beta$)$_{red}^{a}$ & Implied E(B-V)$^b$ & Stellar E(B-V)$^c$ & Maximum Galactic E(B-V)$^d$ &  Literature E(B-V) & Literature E(B-V) Ref.\\
	\hline 
	\textit{Galactic}\\

	Anon    & WR 8     &      & 0.04  & 1.73 & 0.02  & 0.85   & 1.23 &      & \\
	Anon    & WR 16    &      & 0.99  & 2.43 & 0.68  & 0.66   & 2.97 & 0.44 & \citet{M99} \\
	NGC 3199& WR 18    & 1.18 &       &      & 0.81  & 0.94   &      & 0.79 & \citet{K84}       \\
	        &          &      &       &      &       &        &      & 1.10 &  \citet{1992AA...259..629E}   \\
	RCW 58  & WR 40    & 1.07 &       &      & 0.74  & 0.50   & 0.79 & 0.77 & \citet{K84} \\
	\\
	\textit{LMC}\\

	Anon    & BAT99-2  & 0.42 &       &      & 0.29  &    & 0.09 & 0.20 & \citet{2003AA...401L..13N} \\
	Anon    & BAT99-11 &      & 0.16  & 0.62 & 0.11  &    & 0.06 &      & \\  	
	Anon    & BAT99-38 & 0.06 &       &      & 0.04  &    & 0.05 &      & \\ 	
	\hline
	\end{tabular}
	
	\bigskip
	\textit{a}: In some cases different c(H$\beta$) were adopted for blue/red settings, see text. \\
	\textit{b}: Calculated using relations from \citet{1983MNRAS.203..301H} with c(H$\beta$) or c(H$\beta$)$_{blue}$\\
	\textit{c}: \citet{H00} Galactic; unknown LMC\\
	\textit{d}: \citet{1998ApJ...500..525S} Galactic; \citet{1982AJ.....87.1165B} LMC \\
\end{table*}

Due to the relatively low resolution in both the blue and red spectra, blended lines could be a problem in these data. Most seriously, the standard density diagnostic line pairs [S~{\sc ii}] 6717, 6731\AA\ and [O~{\sc ii}] 3727. 3729\AA\ were blended to some degree. The [O~{\sc ii}] doublet is completely unresolved: its measured FWHM is very close to the 8\AA\ instrumental resolution. The [S~{\sc ii}] doublet is marginally resolved by the 14\AA\ resolution in the red spectra, in that there are clearly two components. This situation is far from ideal as this leaves the [S~{\sc ii}] doublet as the only density diagnostic, and with a larger uncertainty on the [S~{\sc ii}] doublet ratio due to its semi-blended nature.

It was also found that the observed intensity ratio of the [O~{\sc iii}] lines of 5007\AA\ to 4959\AA\ was not consistent with theory. This intensity ratio is fixed by the relative atomic transition probabilities and has a value of 2.98 \citep{2000MNRAS.312..813S}, whereas we observed it to be nearer 2.0 in all cases. This was attributed to the 5007\AA\ line falling very close (within 3--4 pixels) to the edge of the detector, and the flux calibration being unreliable there. It was assumed that the 4959\AA\ line was correctly flux calibrated and so it was used on its own at the [O~{\sc iii}] temperature diagnostic stage.

\subsection{UVES Spectra}

The calibrations for UVES observations that are provided by ESO are intended for point source objects, however, and included a sky-subtraction step which was inappropriate for our observations. By default the UVES reduction pipeline assumes that two strips along the edges of the slit are free of the science target spectrum and treats them as sky pixels. The pipeline then averages between these sky pixels and subtracts this from the overall reduced spectrum. The nebulosity of NGC 3199 is very much larger than the $\sim$ 15" slit and as such no region of the slit represented a ``sky" spectrum. The pipeline was therefore re-run with the sky subtraction phase excluded for all the UVES observations.

Cosmic ray subtraction was performed for the UVES observations by median combination of our four science frames for each setting. This produced a final science frame for each setting. The flux calibration was checked by comparing line fluxes in the overlap regions between detectors, which were found to agree to better than $10\%$. 

\begin{figure}
	\centering
	\includegraphics[width=85mm]{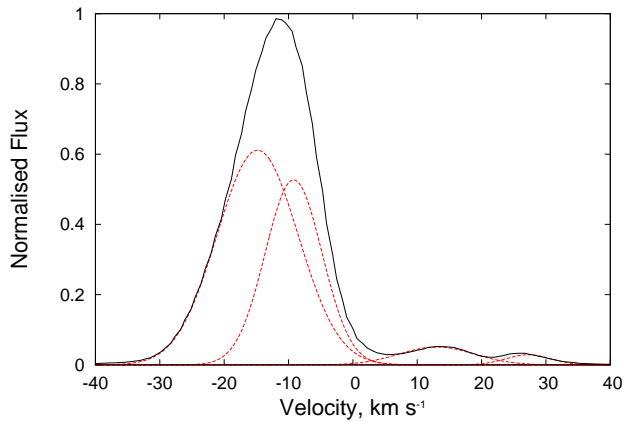}
	\caption{UVES velocity profile of [S~{\sc iii}] 9531\AA\ for NGC 3199, observed profile in black, four-component Dipso/ELF fit dotted.}
	\label{SIII_prof}
\end{figure}

The EFOSC2 problem of line blends was, of course, absent from the UVES observations and lines which are usually blended in lower resolution spectra are well resolved, e.g. [O~{\sc ii}] 7319, 7320, 7330, 7331\AA. In fact the higher resolution of the UVES instrument made the velocity structure of the nebula evident in the bright lines. The best resolution of this velocity structure is seen in the far red [S~{\sc iii}] lines, due to their high atomic weight minimising thermal broadening effects (Figure ~\ref{SIII_prof}). It was found that the best fit to this structure utilised four components, two of which comprised the majority of the emission while a further two are distinctly seen redward of the main emission. 

The lines generated by lower mass ions are subject to a greater degree of thermal broadening, widening the redshifted components to the point where they are indistinguishable. As such it proved difficult to derive a separate reddening solution for the second velocity component. In fact it proved difficult to fit the line profiles with a consistent number of velocity components. The adopted solution was to fit the brighter component with one gaussian, if it produced an accurate fit (the strongest lines: H$\alpha$, [N~{\sc ii}], [O~{\sc iii}] etc) and to use an extra component where it was required and subsequently to sum the components in the bright section.

For NGC 3199 the reddening was derived from the UVES spectrum Balmer decrement. The c(H$\beta$) that we derived, listed in Table~\ref{Redtab} with those of the other nebulae, largely agreed with previous derivations by \citet{K84} and \citet{1992AA...259..629E}. The minor differences in reddening are likely due to the fact that each study has observed a different section of a large nebula.

\section{Results}

\begin{table*}
	\centering                               
	\caption{EFOSC2 Line Intensities (H$\beta$ = 100) - LMC Nebulae }	
	\scalebox{0.9}{
	\begin{tabular}{lccccccccccccc}
	\hline	
	Line & Ion & Observed & $\pm$ & Dereddened & $\pm$  & Observed & $\pm$ & Dereddened & $\pm$ & Observed & $\pm$ & Dereddened & $\pm$ \\
	     &     & \multicolumn{4}{c}{\textbf{BAT99-2}}  & \multicolumn{4}{c}{\textbf{BAT99-11}} & \multicolumn{4}{c}{\textbf{BAT99-38}}\\
	\hline
       3727.0 & [O \textsc{ii}]   &    58.53 &  0.72 &  75.00 &   0.92 &  260.95  &   0.97  & 286.59   &  1.06   &    168.10  &  8.94 &  174.01  &  9.26\\ 
       3868.7 & [Ne \textsc{iii}] &    53.11 &  0.69 &  66.36 &   0.86 &    4.21  &   0.80  &   4.58   &  0.87   &      9.58  &  6.91 &    9.89  &  7.13\\ 
       4101.7 &  H$\delta$     &    24.49 &  1.43 &  29.19 &   1.71 &   23.23  &   0.80  &  24.82   &  0.85   &     24.51  &  0.70 &   25.12  &  0.72\\ 
       4340.5 &  H$\gamma$     &    42.11 &  1.43 &  47.60 &   1.61 &   46.15  &   0.79  &  48.33   &  0.83   &     43.66  &  0.69 &   44.41  &  0.71\\ 
       4363.2 & [O \textsc{iii}]  &    12.85 &  1.42 &  14.45 &   1.60 &    2.76  &   0.79  &   2.88   &  0.82  \\
       4471.5 & He \textsc{i}     &     1.77 &  0.32 &   1.94 &   0.35 &    4.73  &   0.78  &   4.90   &  0.81   &      3.82  &  0.70 &    3.87  &  0.70\\ 
       4685.7 & He \textsc{ii}    &    72.52 &  0.68 &  75.61 &   0.71 \\                                       
       4711.4 & [Ar \textsc{iv}]  &    13.21 &  0.63 &  13.69 &   0.65 \\                                       
       4740.2 & [Ar \textsc{iv}]  &     8.64 &  0.62 &   8.89 &   0.63 \\                                       
       4861.3 &  H$\beta$      &   100.00 &  0.82 & 100.00 &   0.82 &  100.00  &   0.81  & 100.00   &  0.81   &    100.00  &  1.13 &  100.00  &  1.13\\ 
       4958.9 & [O \textsc{iii}]  &   212.87 &  1.48 & 208.00 &   1.45 &  140.47  &   0.83  & 139.25   &  0.82   &     76.92  &  0.74 &   76.67  &  0.74\\ 
       5006.8 & [O \textsc{iii}]  &   592.93 &  1.73 & 572.82 &   1.67 &  352.67  &   0.96  & 348.10   &  0.95   &    178.59  &  0.85 &  177.73  &  0.85\\ 
       6312.1 & [S \textsc{iii}]  &     2.62 &  0.33 &   1.99 &   0.25 &    3.06  &   1.38  &   2.04   &  0.92   &      0.71  &  0.27 &    0.69  &  0.26\\ 
       6562.8 & H$\alpha$      &   394.99 &  1.64 & 289.88 &   1.20 &  449.79  &   3.05  & 284.95   &  1.93   &    290.15  &  2.07 &  277.89  &  1.98\\ 
       6583.5 & [N \textsc{ii}]   &     6.55 &  1.42 &   4.79 &   1.04 &   20.82  &   2.51  &  13.13   &  1.58   &     13.80  &  1.69 &   13.22  &  1.62\\ 
       6678.2 & He \textsc{i}     &          &       &        &        &    6.29  &   2.51  &   3.89   &  1.55   &      3.44  &  1.71 &    3.29  &  1.63\\ 
       6716.4 & [S \textsc{ii}]   &    18.95 &  1.66 &  13.62 &   1.19 &   18.36  &   3.15  &  11.28   &  1.93   &     15.83  &  2.12 &   15.11  &  2.03\\ 
       6730.8 & [S \textsc{ii}]   &    11.29 &  1.67 &   8.10 &   1.20 &   12.93  &   3.12  &   7.92   &  1.91   &     11.66  &  2.12 &   11.13  &  2.02\\
       7005.4 & [Ar \textsc{v}]   &     3.43 &  1.30 &   2.37 &   0.90     \\                                    
       7135.8 & [Ar \textsc{iii}] &    11.16 &  1.39 &   7.61 &   0.95 &   19.45  &   2.51  &  11.05   &  1.42   &      9.72  &  1.67 &    9.21  &  1.58\\
       9068.6 & [S \textsc{iii}]  &    16.10 &  0.77 &   9.14 &   0.43 &   29.56  &   0.71  &  12.84   &  0.31   &     10.98  &  0.77 &   10.15  &  0.71\\
       9530.6 & [S \textsc{iii}]  &    22.26 &  1.33 &  12.25 &   0.73 &   49.91  &   0.88  &  20.69   &  0.36   &      5.03  &  1.15 &    4.62  &  1.06\\
\hline
	\end{tabular}
       }
\label{LineTableLMC}
\end{table*}

 \begin{table*}
	\centering                               
	\caption{EFOSC2 Line Intensities (H$\beta$ = 100) - Galactic Nebulae }	
	\scalebox{0.9}{
	\begin{tabular}{lc|cccc|cccc|cccc}
	\hline
	Line & Ion & Observed & $\pm$ & Dereddened & $\pm$  & Observed & $\pm$ & Dereddened & $\pm$ & Observed & $\pm$ & Dereddened & $\pm$ \\

	     &     & \multicolumn{4}{c}{\textbf{WR 8}}  & \multicolumn{4}{c}{\textbf{WR 16}} & \multicolumn{4}{c}{\textbf{WR 40}}\\
	\hline
         3727.0 &[O \textsc{ii}]    &     60.72  &   6.98  &  62.05  &   7.13 &           &       &        &        &  22.43   &  6.32  &  42.23  &  11.91    \\
         4101.7 &  H$\delta$     &     23.93  &   4.04  &  24.30  &   4.11 &    22.63  &  2.12 &  34.28 &   3.22 &  17.36   &  0.86  &  27.17  &   1.35      \\
         4340.5 &  H$\gamma$     &     48.52  &   4.03  &  49.05  &   4.07 &    28.86  &  2.12 &  38.56 &   2.84 &  35.20   &  0.87  &  48.12  &   1.19        \\
         4471.5 & He \textsc{i}     &  & & & & & &  6.00   &  0.85  &   7.60  &  1.08            \\
	 4861.3 & H$\beta$       &    100.00  &   4.20  & 100.00  &   4.20 &   100.00  &  2.52 & 100.00 &   2.52 & 100.00   &  1.03  & 100.00  &   1.03           \\
         4958.9 & [O \textsc{iii}]  &    109.15  &   4.27  & 108.93  &   4.26 &     7.17  &  2.18 &   6.79 &   2.06 \\                                                
         5006.8 & [O \textsc{iii}]  &    226.13  &   4.84  & 225.45  &   4.82 &     7.67  &  2.12 &   7.07 &   1.95 &   8.82   &  0.86  &   8.08  &   0.79            \\
         6548.1 & [N \textsc{ii}]   &   108.59   &   7.83  &  30.40  &   2.19 &   288.38  & 14.40 &  48.57 &   2.42 & 144.02   &  5.77  &  65.73  &   2.63            \\
         6562.8 & H$\alpha$      &   1026.07  &   9.40  & 284.87  &   2.61 &  1711.12  & 16.04 & 284.82 &   2.67 & 631.23   &  6.23  & 286.59  &   2.83            \\
         6583.5 & [N \textsc{ii}]   &   438.50   &   4.91  & 120.32  &   1.34 &  1140.75  & 12.77 & 186.78 &   2.09 & 543.30   &  5.04  & 244.89  &   2.27            \\
         6678.2 & He \textsc{i}     &      43.01 &   4.69  &  11.19  &   1.22 &    55.69  & 12.10 &   8.46 &   1.84 &  17.91   &  4.67  &   7.81  &   2.03            \\
         6716.4 & [S \textsc{ii}]   &     16.99  &   6.30  &   4.32  &   1.60 &           &       &        &        &  13.82   &  5.51  &   5.95  &   2.37            \\
         6730.8 & [S \textsc{ii}]   &     11.27  &   6.13  &   2.84  &   1.55 &           &       &        &        &   9.92   &  5.47  &   4.25  &   2.34            \\
         7135.8 & [Ar \textsc{iii}] &     47.69  &   4.63  &   9.76  &   0.94 &    43.66  & 10.92 &   4.74 &   1.18 &  13.58   &  4.75  &   5.11  &   1.78             \\
         9068.6 & [S \textsc{iii}]  &     65.30  &   6.31  &   6.28  &   0.60 &           &       &        &        &  19.45   &  1.71  &   4.60  &   0.40            \\
         9530.6 & [S \textsc{iii}]  &     84.69  &   9.40  &   7.14  &   0.79 &           &       &        &      	 &  41.63   &  2.56  &   9.07  &   0.55            \\
\hline
	\end{tabular}
       }

\label{LineTableMW}
\end{table*}

\begin{table}
	\centering
	\caption{UVES Line Intensities (H$\beta$ = 100) - NGC 3199 }	
	\scalebox{0.9}{
	\begin{tabular}{lp{0.9cm}p{1.1cm}p{1.1cm}p{1.1cm}p{1.1cm}}	
	\hline	
	Line & Ion & Observed &$\pm$  & Dereddened & $\pm$  \\
	\hline
        3726.0 &[O {\sc ii}]   &   87.62  &    0.43 &  175.51  &  0.86 \\
        3728.8 &[O {\sc ii}]   &  107.42  &    0.52 &  214.88  &  1.05 \\
        3868.7 &[Ne {\sc iii}] &   36.24  &    0.22 &   67.62  &  0.41 \\
        3967.5 &[Ne {\sc iii}] &   11.16  &    0.19 &   19.72  &  0.34 \\
        4101.7 & H$\delta$     &   15.11  &    0.22 &   24.70  &  0.36 \\
        4340.5 & H$\gamma$     &   32.91  &    0.25 &   46.38  &  0.35 \\
        4363.2 &[O {\sc iii}]  &    2.45  &    0.08 &    3.41  &  0.11 \\
        4471.5 & He {\sc i}    &    3.37  &    0.13 &    4.37  &  0.17 \\
        4861.3 & H$\beta$      &  100.00  &    0.24 &  100.00  &  0.24 \\
        4958.9 &[O {\sc iii}]  &  218.60  &    0.81 &  204.88  &  0.76 \\
        5006.8 &[O {\sc iii}]  &  680.69  &    2.26 &  618.02  &  2.05 \\
        5517.7 &[Cl {\sc iii}] &    1.85  &    0.14 &    1.22  &  0.09 \\
        5537.6 &[Cl {\sc iii}] &    1.40  &    0.08 &    0.91  &  0.05 \\
        5577.3 &[O {\sc i}]    &  0.25$^*$&    0.22 &    0.16  &  0.14 \\
        5875.7 & He {\sc i}    &   21.93  &    0.14 &   12.27  &  0.08 \\
        6300.3 &[O {\sc i}]    &   17.17  &    0.39 &    8.01  &  0.18 \\
        6312.1 &[S {\sc iii}]  &    5.92  &    0.07 &    2.75  &  0.03 \\
        6363.8 &[O {\sc i}]    &    5.78  &    0.14 &    2.62  &  0.06 \\
        6548.1 &[N {\sc ii}]   &   91.97  &    0.40 &   38.90  &  0.17 \\ 
        6562.8 & H$\alpha$     &  675.20  &    1.33 &  283.96  &  0.56 \\
        6583.5 &[N {\sc ii}]   &  287.83  &    1.16 &  120.09  &  0.48 \\
        6678.2 & He {\sc i}    &    8.79  &    0.11 &    3.54  &  0.04 \\
        6716.4 &[S {\sc ii}]   &  118.35  &    0.88 &   46.96  &  0.35 \\
        6730.8 &[S {\sc ii}]   &   96.39  &    0.65 &   38.04  &  0.26 \\
        7135.8 &[Ar {\sc iii}] &   61.93  &    0.37 &   21.19  &  0.12 \\
        7318.9 &[O {\sc ii}]   &    3.06  &    0.07 &    0.99  &  0.02 \\
        7320.0 &[O {\sc ii}]   &   11.40  &    0.08 &    3.68  &  0.02 \\
        7329.7 &[O {\sc ii}]   &    6.07  &    0.07 &    1.95  &  0.02 \\
        7330.7 &[O {\sc ii}]   &    6.15  &    0.07 &    1.98  &  0.02 \\
        8727.1 &[C {\sc i}]    &    1.00  &    0.07 &    0.22  &  0.01 \\
        9068.6 &[S {\sc iii}]  &  480.20  &    5.56 &   98.72  &  1.14 \\
        9530.6 &[S {\sc iii}]  & 1416.72  &   14.39 &  266.44  &  2.70 \\
        9824.1 &[C {\sc i}]    &    3.44  &    0.20 &    0.62  &  0.03 \\
        9850.2 &[C {\sc i}]    &    8.94  &    0.13 &    1.59  &  0.02 \\
	\hline	
	\end{tabular}
	}

      \bigskip
      \textit{*} upper limit
\label{LineTableNGC3199}
\end{table}	                                                                       
 
Observed and dereddened relative line intensity lists for each nebula observed with EFOSC2 can be found in Tables~\ref{LineTableLMC} (LMC) and \ref{LineTableMW} (Galactic). A corresponding line list for NGC 3199 is shown in Table~\ref{LineTableNGC3199} 

The [C~{\sc i}] 8727, 9824, 9850\AA\ lines were not detected in our NTT/EFOSC2 observations. However, in the VLT/UVES observations of NGC 3199 we did detect the [C~{\sc i}] lines, giving an example of what their expected intensity might be. The ratio of the dereddened intensities of [C~{\sc i}] 9850\AA\ to [S~{\sc iii}] 9531\AA\ in NGC 3199 is 0.006. If we apply this ratio to the NTT/EFOSC2 observations we find expected line fluxes of less than half the faintest detected line. We conclude from this that the [C~{\sc i}] lines are not strong enough in any of the NTT/EFOSC2 sample to be detected with the configuration which we utilised, or indeed with any of the available grisms for EFOSC2.

The detection of [C~{\sc i}] 8727, 9824, 9850\AA\ in NGC 3199 is largely due to the high resolution of the UVES instrument. The enhanced resolution did not yield the detection of any heavy element recombination lines. The brightest carbon recombination line (C~{\sc ii} 4267\AA) was not detected despite experimenting with spectral binning of the high resolution observations to attempt to increase the S/N. The relatively high reddening towards NGC 3199 (E(B-V) = 0.81), does not help the detection of this blue line.

\section{Analysis}\label{sec_anal}

\subsection{Plasma Diagnostics}\label{sec_diag}

In total, four different density diagnostic line doublets were employed during the analysis. The [S~{\sc ii}] doublet was present in most cases, however the [O~{\sc ii}], [Ar~{\sc iv}] and [Cl~{\sc iii}] doublets were also used when the instrumental resolution and nebular ionisation permitted. 

All but one of the nebulae (WR 16) displayed the [S~{\sc ii}] doublet at 6716~\AA,6731~\AA, which allowed calculation of their electron densities. Possible reasons for the absence of these lines in the nebula around WR 16 are discussed in Section~\ref{disc_sec}. In each case where the [S~{\sc ii}] doublet was detected, its ratio indicated electron densities near the low density limit  ($1 < n_e < \sim 300$ cm$^{-3}$). In cases where the [S~{\sc ii}] doublet ratio indicated a density less than 50 cm$^{-3}$, a density of 50 cm$^{-3}$ was adopted. As discussed in Section~\ref{DR_sec}, the NTT/EFOSC detections of the [S~{\sc ii}] doublet are blended. This results in a relatively high uncertainty on both the line measurements and subsequently the ratio of the two lines. In contrast the ratio of the [S~{\sc ii}] lines detected in NGC 3199 with UVES carries a much lower uncertainty as both lines were well resolved.

The higher excitation nebulosity around BAT99-2 displayed the [Ar~{\sc iv}] 4711\AA, 4740\AA\ doublet. It yielded a density consistent with the [S~{\sc ii}] density, i.e. the low density limit.

\begin{table*}
  \centering
  \caption{Nebular Diagnostics - NTT/EFOSC Observations}
  \begin{tabular}{l|cccccccc}
	\hline
   Nebula & [S~{\sc ii}]  & n$_e$([S~{\sc ii}]) cm$^{-3}$ &[Ar~{\sc iv}] &n$_e$([Ar~{\sc iv}]) cm$^{-3}$ &[S~{\sc iii}]    & T$_e$([S~{\sc iii}]) K &[O~{\sc iii}]    & T$_e$([O~{\sc iii}]) K  \\ 

 & $ \frac{ 6716 \AA }{ 6731 \AA } $ & & $ \frac{ 4711 \AA }{ 4740 \AA } $ & & $ \frac{ 9069 \AA + 9531 \AA }{ 6312 \AA }^a $  & & $ \frac{ 4959 \AA }{ 4363 \AA } $ & \\
   \hline 
      BAT99-2  & 1.66 $\pm$ 0.29  & 50$^b$  & 0.65 $\pm$ 0.06 & 50$^b$ & 15 $\pm$ 3 & 20000$^{c}$ & 14.4 & 16300 $\pm$ 1000	\\
      BAT99-11 & 1.41 $\pm$ 0.43  & 50$^{+440}_{-50}$ & & & 17$^{+16}_{-4}$ & 20000$^{c}$  & 48 & 10400 $\pm$ 1000 	\\
      BAT99-38 & 1.35 $\pm$ 0.31  & 78$^{+380}_{-77}$& & & 41$^{+32}_{-8}$ & 11200 $\pm$ 2200	\\
	\hline 
 \end{tabular}
    
  \bigskip
  	\textit{$^a$}: [S~{\sc iii}] 9531\AA\ was tellurically absorbed, its intensity was calculated using the ratio I(9531\AA)/I(9069\AA) = 2.48 \\
	\textit{$^b$}: Low density limit.\\
	\textit{$^c$}: High temperature limit.\\
 
\label{Nebular_Diagnostics_EFOSC}
\end{table*}

The increased resolution and wavelength coverage of the UVES instrument allowed the use of the other two line pairs mentioned, [O~{\sc ii}] and [Cl~{\sc iii}]. The densities yielded were consistent with each other and with the [S~{\sc ii}] density (see Table~\ref{NGC_3199_Nebular_Diagnostics}). The [O~{\sc ii}] density in particular agrees very well with the [S~{\sc ii}] density and carries a low uncertainty. The [Cl~{\sc iii}] lines are comparatively weak, so the uncertainty in the ratio is much higher. This results in a less well constrained diagnostic, albeit one that peaks in roughly the same place as the [O~{\sc ii}] and [S~{\sc ii}] densities.

Since the derived electron densities were all below the critical densities of the abundance diagnostic lines, the uncertainties associated with the electron densities have no effect on the abundance uncertainties we will derive.

In general, the electron temperature diagnostics are more important to the final abundance determinations due to the Boltzmann factor dependence of the emissivity of collisionally excited lines. With this in mind, it was of importance to exploit all the available temperature diagnostics. The NTT/EFOSC observations were designed to access the [O~{\sc iii}] 5007, 4959, 4363~\AA\ lines in the blue and the [S~{\sc iii}] 9532, 9069, 6312~\AA\ lines in the red. These diagnostics are strongly dependent upon the very weak, temperature sensitive, [O~{\sc iii}] 4363~\AA\ and [S~{\sc iii}] 6312~\AA\ lines. These lines were detected in some cases, typically where high temperatures might be expected (BAT99-2 for example). The [S~{\sc iii}] 6312~\AA\ line detections, in particular, should be treated as upper limits as they were typically blended with [O~{\sc i}] 6300. 

The VLT/UVES observations yielded a richer set of temperature diagnostics. In addition to the [O~{\sc iii}] and [S~{\sc iii}] lines, which should be regarded as the \textit{primary} temperature diagnostics, temperature diagnostics involving [O~{\sc i}], [O~{\sc ii}] and [C~{\sc i}] were also detected. The UVES instrument, in the dichroic configuration we utilized, has an unfortunate wavelength coverage gap between the red and blue arms of the spectrograph in the region of [N~{\sc ii}] 5755~\AA\ which precluded the use of the [N~{\sc ii}] 6584, 6548, 5755~\AA\ lines as a temperature diagnostic. The [O~{\sc ii}] temperature diagnostic (7319\AA\ + 7320\AA\ + 7330\AA\ + 7331\AA\ versus 3726\AA\ + 3729\AA) was calculated but was ultimately disregarded for two reasons. Firstly, the uncertainty on the diagnostic ratio is rather high due to the weakness of the [O~{\sc ii}] lines at 7325\AA; secondly, this ratio has proven vulnerable to several further systematic uncertainties due to the large wavelength separation between the lines and the possibility of a significant recombination line contribution to the 7320+7330~\AA\ lines (e.g. \citealt{2000MNRAS.312..585L}).

\begin{figure}
	\centering
	\includegraphics[width=85mm]{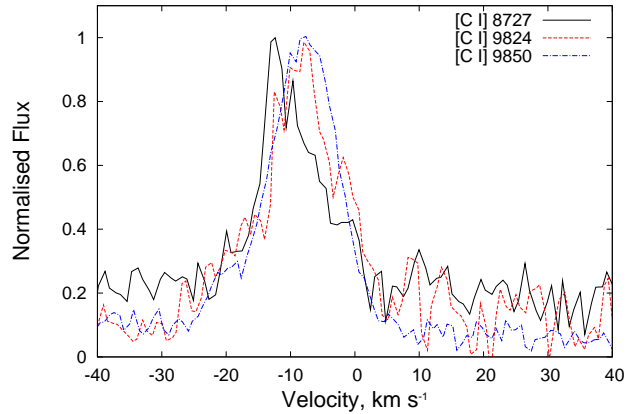}
	\caption{Comparison of the NGC 3199 flux normalised velocity profiles of [C~{\sc i}] 9850\AA\ (dot dash), 9824\AA\ (dashed) and 8727\AA\ (solid). The [C~{\sc i}] 8727\AA\ line appears to have a narrow recombination component.}
	\label{CIprof_fig}
\end{figure}

\subsection{The [C~{\sc i}] far-red lines}

The velocity profiles for the [C~{\sc i}] lines as observed in NGC 3199 are shown in Figure~\ref{CIprof_fig}. The [C~{\sc i}] 8727\AA\ line appears to have two distinct components, one narrow and one of similar width to the other [C~{\sc i}] lines. (a) If we assume that the narrow component is generated by noise and fit the 8727\AA\ line with one component, the [C~{\sc i}] ratio (9850\AA + 9824\AA / 8727\AA) takes a value of around 8--9, which would usually indicate recombination as the origin of the lines. (b) Conversely, if we fit the 8727\AA\ line with a narrow and a wide component, we obtain a [C~{\sc i}] ratio of around 10--11, which is closer to the range of values which would indicate that the lines are collisionally excited in nature. On Figure~\ref{CIprof_fig} we have plotted a box representing the uncertainty range of the [C~{\sc i}] ratio and the likely temperature range under consideration.

\begin{figure}
	\centering
	\includegraphics[width=85mm]{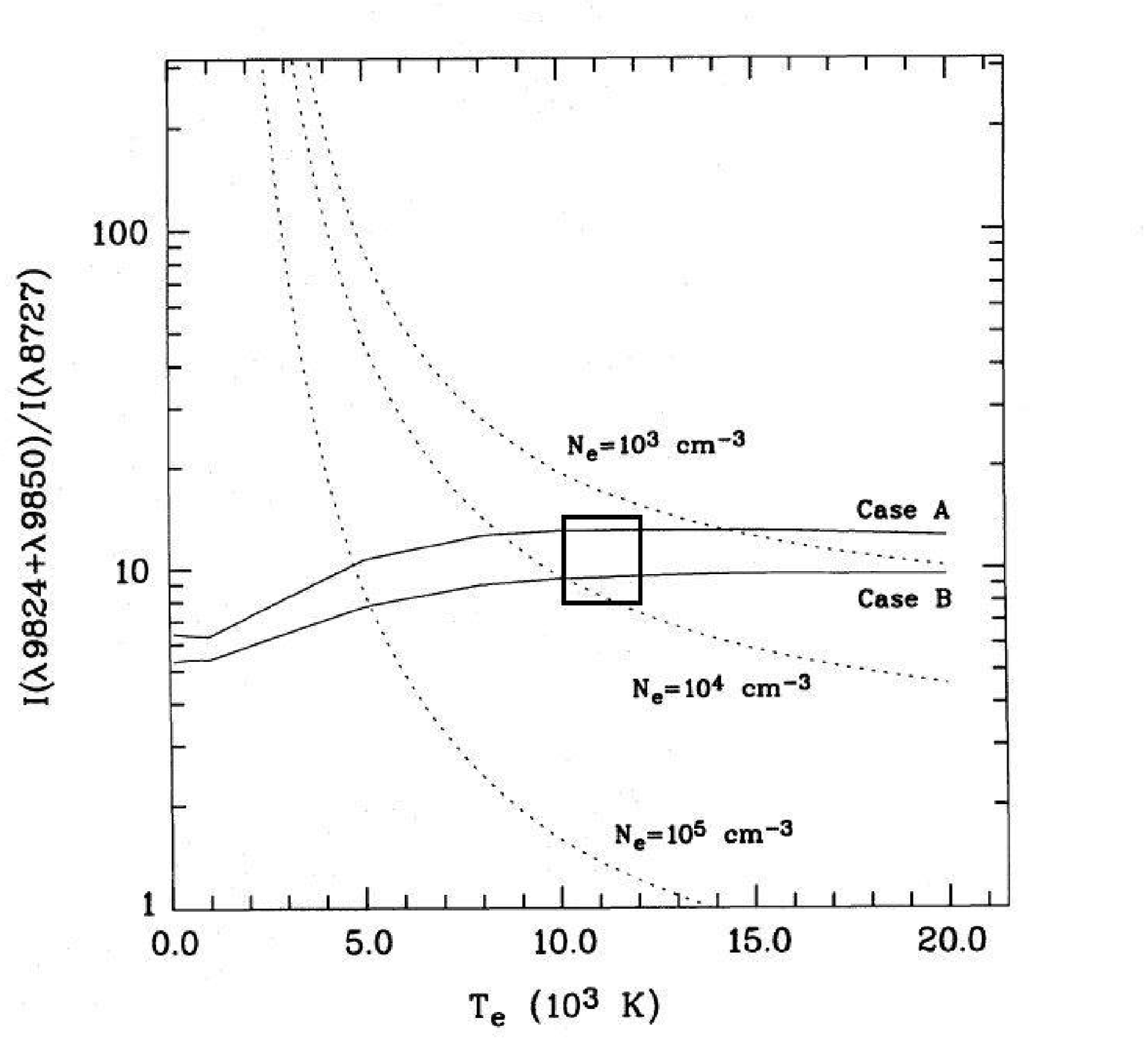}
	\caption{The temperature sensitivity of the [C~{\sc i}] 9850\AA + 9824\AA / 8727\AA\ ratio, in cases where the lines are generated from recombination, (solid lines labelled for Case A \& B recombination respectively, \citealt{1990ApJS...73..513E}) and for cases where the lines are collisionally excited (dotted lines plotted for various electron densities). Reproduced from \citet{1995MNRAS.273...47L}. We have overplotted a box representing the likely margins of error of our [C~{\sc i}] ratio measurement and likely nebular temperatures.}
	\label{LB2005_fig2}
\end{figure}

Figure 2 of \citet{1995MNRAS.273...47L}, reproduced here as Figure~\ref{LB2005_fig2}, shows the sensitivity to electron temperature and density of collisionally excited [C~{\sc i}] 9850\AA\ + 9824\AA / 8727\AA\ (dotted lines). For Case B recombination (indicated by [C~{\sc i}] ratios of 8-9) the temperature dependence is very weak (solid lines) \citep{1990ApJS...73..513E}. It should be noted though that the 9824\AA\ and 9850\AA\ lines are emitted in a fixed intensity ratio (1:2.96), we observe these lines to have a ratio of 2.6 -- suggesting that the uncertainty on the [C~{\sc i}] ratio is a little higher than that quoted. 

From Figure~\ref{LB2005_fig2} we can see that at T=12,000 K the difference between the [C~{\sc i}] diagnostic ratios expected of Case B recombination and collisional excitation (using the lowest density track) is of the order 5-10. Using the quoted line ratio (2.96) we can find the [C~{\sc i}] ratio using either 9850\AA\ or 9824\AA, the maximum [C~{\sc i}] ratio occuring when we use only the 9824\AA\ line -- yielding a value of around 11. The [C~{\sc i}] diagnostic ratio does not therefore unambiguously indicate the mechanism generating the lines.

When the [C~{\sc i}] lines are observed as recombination lines they are usually thought to be emanating from relatively cold, dense, photo-dissociation regions (PDRs) surrounding the ionized regions. \citet{1991ApJ...375..630E} showed that these lines can emanate from PDRs next to high density molecular material, which has been detected in the vicinity of NGC 3199 \citep{2001ApJ...563..875M}. This material could be the origin of the narrow component of the [C~{\sc i}] 8727\AA\ line (while providing a relatively small contribution to the observed 9824 and 9850\AA\ lines). The velocity offset of the peak of the narrow and broad components of [C~{\sc i}] 8727\AA\ is possibly evidence of this.

\begin{figure}
	\centering
	\includegraphics[width=85mm]{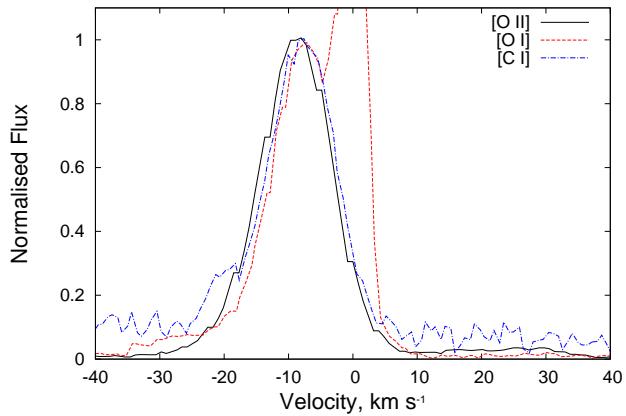}
	\caption{Comparison of the velocity profiles of [O~{\sc i}] 6300\AA\ (dashed), [C~{\sc i}] 9850\AA\ (dot dash) and [O~{\sc ii}] 3726\AA\ (solid) in the UVES spectrum of NGC 3199. The narrow component of the [O~{\sc i}] 6300\AA\ line arises from night sky emission.      }
	\label{CIOIOII_fig}
\end{figure}

In Figure~\ref{CIOIOII_fig} we show the velocity profiles of [O~{\sc i}] 6300\AA, [C~{\sc i}] 9850\AA\ and [O~{\sc ii}] 3726\AA. The [O~{\sc i}] and [O~{\sc ii}] lines are collisionally excited in the nebula, with a characteristic temperature of 8000--10000 K. The line profiles are therefore thermally broadened (which we will use to derive temperatures in the next Section). The fact that the [C~{\sc i}] 9824 and 9850\AA\ lines have very similar widths to those of the [O~{\sc i}] and [O~{\sc ii}] lines indicates that they are generated at similar temperatures.

\begin{table*}  
  \centering
  \caption{NGC 3199 Nebular Diagnostics}
  \begin{tabular}{l|c|cc}
	\hline	
	Ion & Diagnostic & Ratio & Electron Density or Temperature   \\
	\hline	
	{[}S \textsc{ii}]   & $ \frac{ 6716\AA }{ 6731\AA } $                                        & 1.23 $\pm$ 0.01 & 220 $\pm$ 20  cm$^{-3}$ \\
	{[}O \textsc{ii}]   & $ \frac{ 3729\AA }{ 3726\AA } $                                        & 1.22 $\pm$ 0.01 & 230 $\pm$ 20  cm$^{-3}$ \\
	{[}Cl \textsc{iii}] & $ \frac{ 5717\AA }{ 5737\AA } $                                        & 0.744$^{+0.069}_{-0.058}$ & 500 $\pm$ 330 cm$^{-3}$ \\
	{[}S \textsc{iii}]  & $ \frac{ 9069\AA + 9531\AA }{ 6312\AA }$                               & 136  $\pm$ 2    & 7600 $\pm$ 50    K \\ 
	{[}O \textsc{iii}]  & $ \frac{ 5007\AA + 4959\AA }{ 4363\AA }$                               & 241  $\pm$ 7    & 9600 $\pm$ 100   K \\ 
	{[}O \textsc{ii}]   & $ \frac{ 3726\AA + 3729\AA }{ 7319\AA + 7320\AA + 7330\AA + 7331\AA }$ & 45.5 $\pm$ 0.5  & 10350 $\pm$ 50 K \\ 
	{[}O \textsc{i}]    & $ \frac{ 6300\AA + 6364\AA }{ 5577\AA }$                               & $>$ 51.5     & $<$ 12250 K \\
	{[}S \textsc{ii}]   & $ \frac{ 6717\AA + 6731\AA  }{ 4076\AA + 4068\AA }$                    & 13.9 $\pm$ 0.5  &  9000 $\pm$ 1000   K \\
	{[}C \textsc{i}]    & $ \frac{ 9850\AA + 9824\AA }{ 8727\AA }$                               &  9.8 $\pm$ 0.6  &   \\
	\hline 
	\end{tabular} 
\label{NGC_3199_Nebular_Diagnostics}

\end{table*}

\subsection{Line Broadening Temperatures}

\begin{figure}
	\centering
	\includegraphics[width=85mm]{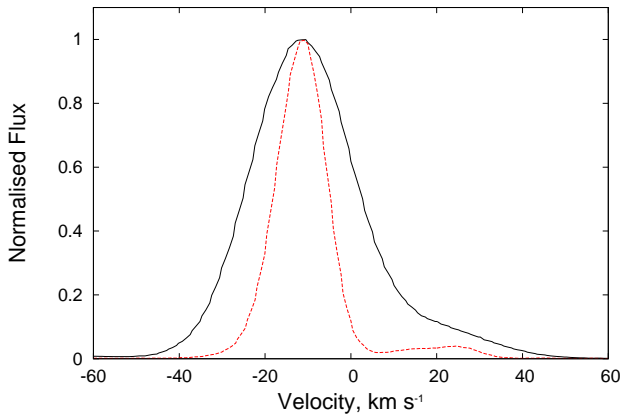}
	\caption{Observed velocity profiles of H$\alpha$ (solid) and [N~{\sc ii}] 6584\AA\ (dashed) in NGC 3199. The H$\alpha$ line is broadened by fine structure components along with the thermal and turbulent broadening that also affect [N~{\sc ii}]. (Flux coordinate has been normalised to allow better comparison.)}
	\label{HA_NII_comp}
\end{figure}

The observed width of a spectral line depends upon several factors, primarily the nebular temperature, turbulence and expansion along with the instrumental line profile. Only the temperature broadening is a function of atomic mass, so if we can measure the widths of several lines from ions of different atomic weight to sufficient accuracy, it is possible to derive the temperature of the line-emitting plasma. Historically this has usually been performed using only a single pair of lines, [N~{\sc ii}] 6584\AA\ and H$\alpha$, due to both their proximity and strengths (e.g. \citealt{1972A&A....17..165D}). The velocity profiles of [N~{\sc ii}] 6584\AA\ and H$\alpha$ in our UVES observations are shown in Figure~\ref{HA_NII_comp}. Given the resolution of the VLT/UVES observations it is possible to accurately determine the widths of a large number of strong emission lines (including the traditional pair) and to derive temperatures from more than one pair.

The major obstacle which must be overcome to perform this process is a detailed description of the intrinsic line shapes. For collisionally excited lines this is relatively simple: most have negligible intrinsic fine structure broadening and all broadening in excess of the instrumental broadening must be a combination of thermal and turbulent broadening. The situation is rather different for recombination lines, which are often comprised of several very closely spaced fine-structure components. The first work done to describe the fine structure components of the H$\alpha$ line in this context found that the ``natural" profile of the line could have a significant effect on the derived electron temperature \citep{1971A&A....12..219D}. In our analysis we shall use the fine structure correction term provided by \citet{1999A&AS..135..359C} [Equation 9] for H$\alpha$, which follows the same form as those derived from earlier formulations while introducing a correction factor $\delta$ which combines with the H$\alpha$ FWHM in quadrature :

\begin{equation}  \label{eq_line_broad} 
T_e = \frac{ \mathrm{FWHM(H\alpha)}^2 - \delta^2 - \mathrm{FWHM( [N~\textsc{ii}])}^2}{ 8k [ \frac{1}{m_H} - \frac{1}{m_N} ] \times ln(2) }
\end{equation}

where $T_e$ is the electron temperature, FWHM(H$\alpha$) and FWHM(N~{\sc ii}) are the FWHMs of H$\alpha$ and [N~{\sc ii}], $m_H$ and $m_N$ are the atomic weights of hydrogen and nitrogen respectively and $k$ is the Boltzmann constant. An explicit equation for calculating $\delta$ is not quoted, but a table of values suitable for interpolation was included by \citet{1999A&AS..135..359C}.

We can generalise Equation~\ref{eq_line_broad} to apply it to other species, replacing FWHM([N~\textsc{ii}]) and $m_N$ with values appropriate for the line and species we wish to use.  

\begin{equation}  \label{eq_line_broad_2} 
T_e = \frac{ \mathrm{FWHM (H\alpha)}^2 - \delta^2 - \mathrm{FWHM ([X])}^2}{ 8k [ \frac{1}{m_H} - \frac{1}{m_X} ] \times ln(2) }
\end{equation}


The results of applying Equation~\ref{eq_line_broad_2} for the strongest collisionally excited lines with H$\alpha$ appear in Table~\ref{T_LB}. Singlet He~{\sc i} lines can be included as they have no fine structure components. The only singlet He~{\sc i} line found with sufficient strength to measure its width accurately was He~{\sc i} 6678.2~\AA. We have not included temperatures derived from [C~{\sc i}] or [O~{\sc i}] in the plots as they are potentially unreliable. Equation~\ref{eq_line_broad_2} is only valid if we believe the lines are being generated by the same gas. In the case of [C~{\sc i}] and [O~{\sc i}], this may not be the case as they are likely to be generated in the outer regions of the nebula.

\begin{table}
	\centering
	\caption{Temperature determinations from line broadening in NGC 3199}
	\begin{tabular}{ccccc}
	\hline	
Species & Line (\AA) &  FWHM (km s$^{-1}$)  & $T_e$ & $\pm$ \\
	\hline
	H{\sc i} 	 & 6562.80& 27.113 $\pm$ 0.045  & -- & -- \\
	{[}N~{\sc ii}]   & 6548.10& 13.411 $\pm$ 0.055  & 11580 & 72 \\
	{[}N~{\sc ii}]   & 6583.50& 13.505 $\pm$ 0.050  & 11520 & 66 \\   
	{[}O~{\sc ii}]   & 3726.03& 13.250 $\pm$ 0.062  & 11570 & 81 \\
	{[}O~{\sc ii}]   & 3728.82& 13.373 $\pm$ 0.062  & 11490 & 80 \\
	{[}S~{\sc ii}]   & 6716.44& 12.629 $\pm$ 0.087  & 11560 & 120 \\
	{[}S~{\sc ii}]   & 6730.82& 12.597 $\pm$ 0.080  & 11570 & 110 \\
	{[}O~{\sc iii}]  & 4958.91& 15.975 $\pm$ 0.054  & 9730  & 52 \\
	{[}O~{\sc iii}]  & 5006.84& 15.840 $\pm$ 0.047  & 9820  & 47 \\
	{[}S~{\sc iii}]  & 9068.90& 14.250 $\pm$ 0.142  & 10570 & 151 \\
	{[}S~{\sc iii}]  & 9531.00& 14.695 $\pm$ 0.175  & 10290 & 175 \\
	{[}Ne~{\sc iii}] & 3869.06& 15.036 $\pm$ 0.086  & 10260 & 86 \\
	{[}Ne~{\sc iii}] & 3967.79& 14.578 $\pm$ 0.234  & 10570 & 241\\
	{[}Ar~{\sc iii}] & 7135.90& 14.267 $\pm$ 0.075  & 10510 & 82\\	
	He~{\sc i}       & 6678.15& 18.739 $\pm$ 0.219  & 9380  & 157 \\	
	{[}C~{\sc i}]    & 9850.26& 12.152 $\pm$ 0.230  & 12490 & 330 \\
	{[}O~{\sc i}]    & 6300.30& 12.175 $\pm$ 0.345  & 12480 & 510 \\
	\hline
	\end{tabular}
 \label{T_LB}	
\end{table}

\begin{figure}
	\centering
	\includegraphics[width=85mm]{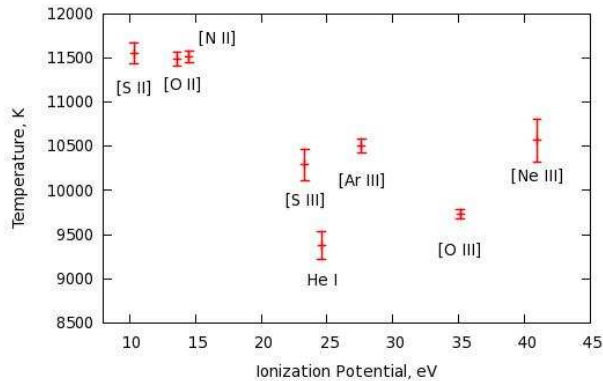}
	\caption{NGC 3199 line broadening temperatures for various ions relative to H$\alpha$, versus the ionization potential to obtain the parent ion.}
	\label{LB_ion_plot}
\end{figure}

In Figure~\ref{LB_ion_plot}, we show the line broadening temperature versus the ionization potential for each species. The species with higher ionisation potentials generally have lower temperatures by at least 1000 K. This trend could be attributable to an effect known as ``radiation hardening" \citep[see][Section 7.3.2]{2005pcim.book.....T}. The ionization cross section of H~\textsc{i} is proportional to $\nu^{-3}$, meaning that  the lowest energy photons are absorbed by the inner regions of the nebula. Hence as radius increases the average energy and heating effect of the remaining ionizing photons also increases. For a radiation bounded nebula, these outer regions will also be where the degree of ionization decreases.

\begin{figure}
	\centering
	\includegraphics[width=85mm]{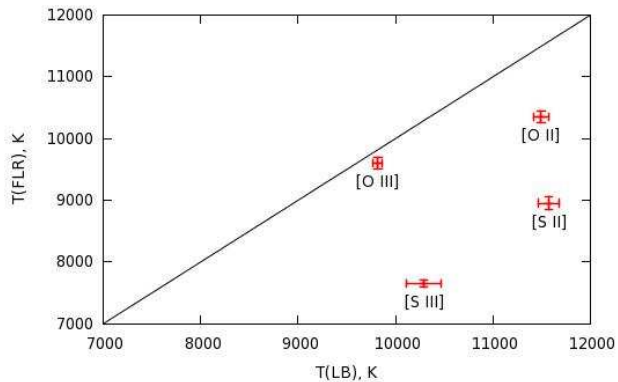}
	\caption{Comparison of line broadening temperatures T(LB) with forbidden line ratio temperatures T(FLR). The black solid line corresponds to T(LB) = T(FLR). }
	\label{LB_temp_plot_comp}
\end{figure} 

Figure~\ref{LB_temp_plot_comp} shows the comparison between temperatures calculated using the line broadening method with those determined using diagnostic ratios. In most cases the line broadening temperature is higher than the diagnostic ratio temperature, [O~{\sc iii}] showing the closest agreement.

\subsection{Nebular Abundances}

For each nebula, diagnostic uncertainties were computed via a novel Monte-Carlo technique. The ``{\sc NEAT}" (Nebular Empirical Analysis Tool) software, developed at UCL, allows an exploration of the parameter space defined by the uncertainties associated with nebular line measurements. A full description of this code and its capabilities will be forthcoming (Wesson et al., 2011, in prep). Briefly, {\sc NEAT} reads in a list of measured spectral lines  (their wavelengths, relative intensities and uncertainties), and generates a ``possible" line list by randomly offsetting the line intensities using the quoted uncertainties as limits. It then performs a full empirical analysis on this ``possible" line list, calculating all diagnostic ratios and plasma parameters, along with a full abundance analysis using a three-zone ionization model by default. It repeats this process as many times as the user requires, building up a picture of the uncertainties associated with each diagnostic, abundance, etc, as the number of data points increases. One then bins these values to obtain an uncertainty distribution. {\sc NEAT} was used to generate all of the values and uncertainties in Table~\ref{NGC_3199_Nebular_Diagnostics}, along with the abundance derivations described later.

The {\sc NEAT} code uses the same statistical equilibrium methods as the earlier {\sc EQUIB} code (developed at UCL by S. Adams and I. Howarth) to derive abundances for CELs, solving the equation:
\begin{equation}
\frac{N(ion)}{N(H^+)} \times A = \frac {I(line)}{I(H\beta)},
\end{equation}

where A is related to the emissivities of the two ions, via the following relationship:

\begin{equation}\label{abund_eq}
 A \propto \frac{\epsilon(line)}{\epsilon(H\beta)} \propto \frac{ n(ion) n_e e^{-\frac{E}{kT_e}}  }{ N(H^+) n_e T^{-0.9}_{e} }
\end{equation}

The A value for the specific ion at given $n_e$ and $T_e$ is calculated and then used to derive $\frac{N(ion)}{N(H^+)}$, the linear abundance of the specific ion as indicated by the particular line chosen. {\sc NEAT} derives uncertainties on the abundance determinations in the same manner as described for the temperature and density diagnostics. The provenance of the atomic data used is shown in Table~\ref{AD_reftable}.

\begin{table}
	\centering
	\caption{Sources of Atomic data}
	\begin{tabular}{p{0.8cm}|p{6.5cm}}
	\hline
	Ion &  Reference\\
	\hline
	C$^0$      &    \citet{1976AA....50..141P}, \citet{1987JPhB...20.2553J}, \citet{1979AA....72..129N} \\ 
	Cl$^{++}$  &    \citet{1982MNRAS.198..127M}, \citet{1989AA...208..337B},  \\
	O$^0$      &    \citet{1981PSS...29..377B}, \citet{1988JPhB...21.1083B}, \citet{1988JPhB...21.1455B} \\
	O$^{+}$   &    \citet{1976MNRAS.177...31P}, \citet{1982MNRAS.198..111Z}\\
	S$^{++}$   &    \citet{1982MNRAS.199.1025M}, \citet{1983IAUS..103..143M}	\\
	All others &	\citet{2006ApJS..162..261L} \\
	\hline
	\end{tabular}
 \label{AD_reftable}	
\end{table}

In the simplest cases, only one of each type of diagnostic was available per nebula e.g. BAT99-38, where T$_e$([S~{\sc iii}]) and $n_e$([S~{\sc ii}]) were available. This led to a very simple abundance determination where the same temperature and density was used for every ion. This was the situation for each of the three LMC WR nebulae. 

For NGC 3199, multiple density and temperature diagnostics were available. Each density diagnostic was found to place the nebula at the diagnostic's low density limit. The forbidden line ratio temperature diagnostics indicated that a multi-zone ionisation model may be appropriate, in that lower ionisation species (e.g. [O~{\sc ii}] and [S~{\sc ii}]) gave higher temperatures than those from moderate ionisation species (e.g. [O~{\sc iii}] and [S~{\sc iii}]). This trend was corroborated by the line broadening temperature measurements (see Figure~\ref{LB_ion_plot}). Ultimately it was decided to use a two zone model where we used the average of the [O~{\sc ii}] and [S~{\sc ii}] forbidden line ratio temperatures for the low ionisation species, the [S~{\sc~iii}] temperature for [S~{\sc iii}], and the [O~{\sc iii}] temperature for all other ions. 

The results of the abundance calculations for the LMC WR nebulae and NGC 3199 are listed in Table~\ref{Total_Abundances} using ICFs (Ionisation Correction Factors) from \citet{1994MNRAS.271..257K}. In Table~\ref{Table_NO} we list the N/O abundance ratio in each of our nebulae along with some comparison objects.

The remaining three Galactic nebulae in our NTT/EFOSC2 sample have no observed temperature diagnostic. As discussed earlier in this Section, the abundances from CELs depend on a Boltzmann factor and as such are sensitive to the adopted electron temperature.

Previous authors have adopted temperatures for some of these objects based on matching the oxygen abundance to those of H~{\sc ii} regions in the same galaxy. However we do not think this approach is appropriate in this case, as the oxygen abundance in WR nebulae could easily be depleted relative to Galactic H~{\sc ii} regions due to CNO cycling by the massive star prior to it becoming a WR star. One solution might be to assume a nebular temperature that yields Galactic H~{\sc ii} region neon abundances, as neon should not have been astrated by stellar processing. Unfortunately the higher reddening in the Galactic plane meant that we had no detections of [Ne~{\sc iii}] 3868, 3967\AA\ for RCW 58 or the nebulae around WR 8 and WR 16, and hence no neon abundance for these objects.

However we can infer that the abundance pattern in these nebulae is not the same as in H~{\sc ii} regions, or indeed in the other WR nebulae in our sample, by simple inspection of the detected lines. WR 16 and RCW 58 in particular display [N~{\sc ii}] lines comparable in intensity to H$\alpha$, indicating likely nitrogen enhancement. 

In the absence of a temperature diagnostic for these nebulae, we employed a different approach to finding one important diagnostic - log(N/O). We know, from our observations and from previous authors, that WR nebulae are likely to possess electron temperatures in the range 7000--12000 K.

The {\sc EQUIB} code, as mentioned previously, can be used to generate the appropriate line emissivities $\epsilon$ across this range (using the atomic data listed in Table~\ref{AD_reftable}). We can then plot N$^+$/O$^+$ as a function of electron temperature T$_e$ as:

\begin{equation}
log\frac{N^+}{O^+} (T_e) = \frac{ \mathrm{I([N~\textsc{ii}] 6584\AA)}  }{ \mathrm{I([O~\textsc{ii}] 3727 + 3729\AA)  }} \times E
\end{equation}

 where:

\begin{equation}
E = \frac {\mathrm{\epsilon_{[O~\textsc{ii}] 3727\AA}(T_e) + \epsilon_{[O~\textsc{ii}] 3729\AA}(T_e) } }{\mathrm{\epsilon_{[N~\textsc{ii}] 6584\AA}(T_e)}}
\end{equation}

\begin{figure}
	\centering
	\includegraphics[width=85mm]{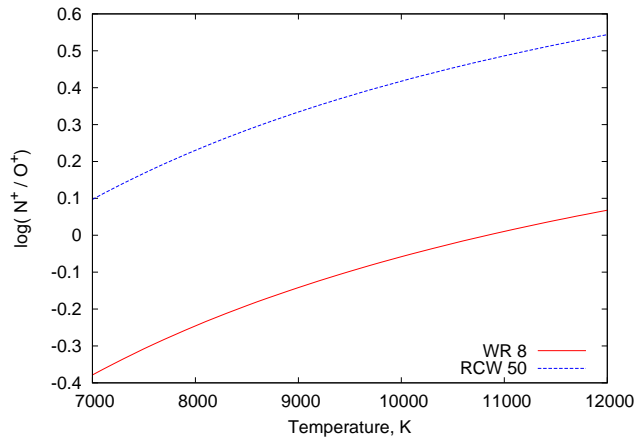}	
	\caption{Range of possible log$\frac{N^+}{O^+}$ values over the range 7000-12000 K for the nebulae around WR 8 and WR 40 (RCW 58). }
	\label{NO_range_plot}
\end{figure}

Where I(line) is the line intensity relative to H$\beta$ = 100 and $\epsilon(T_e)$ is the line emissivity at temperature $T_e$. This $N^+/O^+$ relationship is shown in Figure~\ref{NO_range_plot} for RCW 58 and the nebula around WR 8. As expected, the ratio is not a strong function of temperature and therefore provides a reasonable constraint upon the value of $log\frac{N^+}{O^+}$ and hence the degree of nitrogen enrichment within each nebula. 

As mentioned earlier, we used the icf scheme of \citet{1994MNRAS.271..257K}, which specifies identical icfs for N$^+$ and O$^+$. This means that the N$^+$/O$^+$ ratio that we calculate should be the same as the overall N/O ratio. The range of N/O ratios derived for RCW 58 and for the nebula around WR 8 will be discussed in Section~\ref{disc_sec}. 

\begin{table*} 
  \centering
  \caption{Total Abundances}
  \begin{tabular}{l|ccccccc}
 \hline
                                       & BAT99-2                   & BAT99-11                & BAT99-38 & LMC H\textsc{ii} Region$^a$ & NGC 3199  & Milky Way H\textsc{ii} region$^a$ \\
\hline  
    He$^+$ / H$^+$                     &  0.042 $\pm$ 0.007        & 0.102 $\pm$ 0.015       & 0.083 $\pm$ 0.014       &       & 0.088 $\pm$ 0.006&  \\
    He$^{++}$ / H$^+$                  &  0.066 $\pm$ 0.002 \\   
    12 + log He / H                    &  11.04 $\pm$ 0.02         & 11.01$^{+0.07}_{-0.10}$ & 10.92$^{+0.09}_{-0.12}$ & 10.98 & 10.95 $\pm$ 0.01 & 10.98  \\
    \\

    C$^0$ / H$^+$ ($\times 10^7$)      &                           &                         &                         &       & 5.97 $\pm$ 0.20 &  \\   
    12 + log C$^0$/H$^+$               &                           &                         &                         &       & 5.78 $\pm$ 0.20 &  \\
    \\

    N$^+$ / H$^+$ ($\times 10^6$)      &  0.80$^{+0.15}_{-0.10}$   & 2.28$^{+0.23}_{-0.28}$  & 2.25 $\pm$ 0.20          &       & 21.8 $\pm$ 0.6 \\ 
    N ICF$^b$                          &  4.65$^{+0.50}_{-0.30}$   & 1.53$^{+0.25}_{-0.07}$  & 1.45$^{+0.50}_{-0.05}$  & 2.31  & 2.67 $\pm$ 0.02  &  3.87 \\ 
    12 + log N / H                     &  6.61$^{+0.07}_{-0.11}$   & 6.56 $\pm$ 0.06         & 6.54$^{+0.14}_{-0.06}$  & 6.92  & 7.77 $\pm$ 0.01  &  7.49 \\
    \\

    O$^0$ / H$^+$ ($\times 10^5$)      &                           &                         &                         &       & 1.86 $\pm$ 0.09  &\\
    O$^+$ / H$^+$ ($\times 10^5$)      &  2.80$^{+0.04}_{-0.05}$   & 10.70$^{+0.60}_{-0.20}$ & 6.58$^{+0.57}_{-0.37}$  &       & 17.3 $\pm$ 0.8   &\\
    O$^{++}$ / H$^+$ ($\times 10^5$)   &  4.10$^{+0.40}_{-0.30}$   & 5.90$^{+2.7}_{-0.07}$   & 3.20$^{+4.00}_{-0.85}$  &       & 28.9 $\pm$ 0.8   &\\    
    O ICF$^b$                          &  1.83$^{+0.17}_{-0.08}$   & 1                       & 1                       &       & 1               &  1 \\
    12 + log O / H                     &  8.12 $\pm$ 0.03           & 8.23$^{+0.06}_{-0.02}$  & 8.01$^{+0.14}_{-0.05}$  & 8.41  & 8.66 $\pm$ 0.02 &  8.55  \\
    12 + log O$^0$/H$^+$               &                           &                         &                         &       & 7.28 $\pm$ 0.02 &   \\
    \\    
    log  N / O                         &  -1.52$^{+0.07}_{-0.10}$  & -1.67$^{+0.08}_{-0.12}$ & -1.48 $\pm$ 0.05        & -1.49 & -0.90 $\pm$ 0.01 &  -0.91 \\
    \\	   

    Ne$^{++}$ / H$^+$ ($\times 10^5$)  &  1.13$^{+0.13}_{-0.10}$   & 0.20$^{+0.12}_{-0.03}$  & 0.44$^{+1.10}_{-0.25}$  &       & 10.1 $\pm$ 1.0  &  \\   
    Ne ICF$^b$                         &  3.07$^{+0.33}_{-0.20}$   & 2.60 $\pm$ 0.88         & 2.60 $\pm$ 0.96         & 1.76  & 1.60 $\pm$ 0.01 &  1.39 \\
    12 + log Ne / H                    &  7.55$^{+0.04}_{-0.03}$   & 6.83$^{+0.10}_{-0.11}$  & 7.39$^{+0.24}_{-0.40}$  & 7.62  & 8.23 $\pm$ 0.02 & 7.99 \\
    \\

    S$^{+}$ / H$^+$   ($\times 10^6$)  &  0.50 $\pm$ 0.04           & 0.45$^{+0.07}_{-0.06}$  & 0.59$^{+0.08}_{-0.06}$  &       & 2.10 $\pm$ 0.06 &  \\
    S$^{++}$ / H$^+$  ($\times 10^6$)  &  0.46$^{+0.04}_{-0.03}$    & 1.07$^{+0.25}_{-0.09}$  & 1.02$^{+0.50}_{-0.25}$  &       & 15.6 $\pm$ 0.50 &  \\ 
    S ICF$^b$                          &  1.25 $\pm$ 0.02           & 1.02$^{+0.01}_{-0.01}$  & 1.01$^{+0.02}_{-0.01}$  &       & 1.10 $\pm$ 0.01 &  1.19\\
    12 + log S / H                     &  6.08 $\pm$ 0.02           & 6.20$^{+0.06}_{-0.03}$  & 6.21$^{+0.13}_{-0.07}$  & 6.70  & 7.29 $\pm$ 0.01 &  7.00\\ 
    \\

    Ar$^{++}$ / H$^+$ ($\times 10^6$)  &  0.22$^{+0.03}_{-0.01}$   & 0.62$^{+0.20}_{-0.09}$  & 0.51$^{+0.42}_{-0.11}$  &       & 2.37 $\pm$ 0.05 & \\
    Ar$^{3+}$ / H$^+$ ($\times 10^6$)  &  0.68 $\pm$ 0.01     \\
    Ar$^{4+}$ / H$^+$ ($\times 10^6$)  &  0.14 $\pm$ 0.04  \\
    Ar ICF$^b$                         &  1                        & 1.87                    & 1.87                    & 1.79  & 1.87            &  1.35 \\ 
    12 + log Ar / H                    &  6.02$^{+0.04}_{-0.03}$   & 6.08$^{+0.11}_{-0.07}$  & 6.06$^{+0.21}_{-0.14}$  & 6.30  & 6.65 $\pm$ 0.01 &  6.31\\ 
	\hline
     \end{tabular}

      \bigskip
      \textit{a}: \citet{2003MNRAS.338..687T} (LMC N11B), (M 17)\\
      \textit{b}: Using the Ionisation Correction Factor scheme of \citet{1994MNRAS.271..257K}
 \label{Total_Abundances}
\end{table*}

\begin{table} 
  \centering
  \caption{N/O ratios}
  \begin{tabular}{l|c}
	\hline	
	Nebula   & log(N/O)  \\
	\hline       
	BAT99-2  &  -1.52$^{+0.07}_{-0.10}$  \\
	BAT99-11 &  -1.67$^{+0.08}_{-0.12}$  \\
	BAT99-38 &  -1.48 $\pm$ 0.05   \\
	LMC H~{\sc ii} region & -1.49 \\	
	\hline
	WR 8     &  $>$ -0.4  \\
	WR 16    &  $>$ -0.1  \\
	NGC 3199 &  -0.90 $\pm$ 0.01   \\
	RCW 58   &  $>$ 0.1  \\
	Galactic H~{\sc ii} region & -0.91\\
	Solar$^a$ & -0.86 \\
\hline
  \end{tabular}

\bigskip
$^a$: \citet{2009ARAA..47..481A}

\label{Table_NO}
\end{table}

\section{Notes on Individual Objects}\label{disc_sec}

\subsection*{BAT99-2}

The EFOSC2 optical spectrum of the nebula around BAT99-2 (WN2) is in good agreement with that presented by \citet{2003AA...401L..13N}. The nebula is highly ionized which manifests itself via the presence of [Ar~{\sc iv}] lines along with He~{\sc ii} 4686 \AA, which was only detected in BAT99-2.

The nebula around BAT99-2 does not display any significant under- or over- abundances relative to LMC H~{\sc ii} regions (see Table~\ref{Total_Abundances}).  In general the abundances we derive agree with those derived by \citet{2003AA...401L..13N} and we agree with their conclusion that BAT99-2 does not contain any abundance anomalies.

\subsection*{BAT99-11}

The nebula encircling BAT99-11 (WC4) displays a lower degree of ionization which is more typical of WR nebulae despite having a relatively high excitation (WC4) central star. The [Ne~{\sc iii}] and [Ar~{\sc iv}] lines seen in the nebula around BAT99-2 are not observed. Instead we see pronounced higher order Balmer series lines along with several He~{\sc i} lines, most prominently He~{\sc i} 4471\AA. 

The nebular abundances display no signs of stellar processing relative to H~{\sc ii} region LMC N11B (see Table~\ref{Total_Abundances}). Since BAT99-11 is a WC4 star it had been anticipated that the nebula might contain an ejecta component. Based on the lack of abundance anomalies, the nebula around BAT99-11 is a wind-blown shell.

\subsection*{BAT99-38}

The spectrum of the nebula near BAT99-38 (WC4+O) is very similar to that of the nebula around BAT99-11, albeit roughly a factor of five fainter in surface brightness. Many He~{\sc i} lines are present, including 7065\AA, 6678\AA, 4471\AA\ and several higher order Balmer lines are present blue-wards of 4000\AA. 

The nebula near BAT99-38 displays no abundance enhancements relative to H~{\sc ii} region LMC N11 (see Table~\ref{Total_Abundances}). Based on our results and the fact that the star is superposed on the arc of nebulosity (Figure~\ref{B38_slit}), it is dubious whether the nebula near BAT99-38 is actually related to BAT99-38 at all. 

\subsection*{WR 8}

The nebulosity surrounding WR 8 (WC7/WN4) presents clear evidence of nitrogen enrichment in the form of strong [N~{\sc ii}] 6548, 6584\AA\ emission lines relative to H$\alpha$. There is also some evidence of WR features scattered into the slit, since a wide bump at around 4650\AA\ corresponds to a known WR emission feature. 

From Figure~\ref{NO_range_plot} we can see that the range of log(N/O) indicated, -0.4 $<$ log(N/O) $<$ 0.1, is well above the N/O ratio found for the M17 Galactic  H~{\sc ii} region (-0.91). Even the lower bound, -0.4, is a factor of three above the Galactic ISM value, indicating a degree of nitrogen enhancement relative to oxygen. This supports the conclusion of \citet{2010MNRAS.409.1429S} that this nebula has a significant stellar ejecta component.

\subsection*{WR 16}

The nebulosity around WR 16 (WN8) displays a sparse, but interesting spectrum. We noted earlier that previous observations \citep{M99} had not detected the [S~{\sc ii}] doublet, a result repeated in our spectrum. The spectrum also contains the strongest [N~{\sc ii}] lines in our sample along with no trace of [O~{\sc iii}], [O~{\sc ii}] or [O~{\sc i}] lines, making a direct N/O determination impossible but implying a high N/O ratio. This kind of composition is precisely what one might expect from stellar outflows of a WN type WR star. There is a strong sign of helium overabundance, with He~{\sc i} 6678\AA\ present and stronger than in all but the nebula around WR 8. Given the non-detection of the [O~{\sc ii}] doublet at 3727\AA\ we cannot employ our `bootstrapping" approach to calculate an approximate N/O value for the nebulosity, except to say that it must be strongly nitrogen enriched since if we assume a conservative upper limit for the [O~{\sc ii}] 3727\AA\  intensity, for example $\frac{[O \textsc{ii}] 3727\AA}{H\beta} \sim 0.5$ (as is the case for WR 8 and RCW 58), then we find -0.1 $<$ log(N/O) $<$ 0.3 for the temperature range 7000--12000 K, a factor of 10 higher than the solar N/O value.

\citet{M99} suggested that the lack of detectable [S~{\sc ii}] and [O~{\sc iii}] lines in the spectrum of the WR 16 nebula was evidence of an extremely high nebular density ($>$ 10000 cm$^{-3}$). The critical densities of [S~{\sc ii}] and [O~{\sc iii}] do indeed lie in this range, however we attribute the weakness of the [O \textsc{iii}] lines to the low effective temperature of the WN8 central star.  

\subsection*{NGC 3199 (WR 18)}

NGC 3199 displays very similar abundances to the Galactic H~{\sc ii} region M17 (see Table~\ref{Total_Abundances}). We conclude that the nebulosity is mainly a swept-up ISM shell. Previous observations of this nebula have derived an expansion velocity of around 15~km~s$^{-1}$ \citep{1982ApJ...254..578C,2001ApJ...563..875M}. This value is similar to the velocity separation of the main nebular emission components in our UVES spectrum (see Figure~\ref{SIII_prof}). However, the low intensity of the red-shifted component suggests that it is unlikely to have the same intrinsic luminosity as the blue-shifted component, as the mass of internal dust required to produce the required extinction (A(H$\alpha$) $>$ 4mags) is unphysical. By inspection of Figure~\ref{NGC3199_slit} it is obvious that NGC 3199 is not spherically symmetric, its shape suggesting a swept up origin, in which case the relative intensities and velocities of the emission components most likely are due to variations in the local ISM density.

Following the discussion in Section~\ref{sec_diag}, the dominant excitation mechanism for the observed [C~{\sc i}] lines in NGC 3199 appears to be collisional excitation of C$^0$ in the outer regions of the nebula. The C$^0$/O$^0$ abundance derived, log(C$^0$/O$^0$) = -1.49, would require detailed photoionization modelling to convert into log(C/O) as the physical and ionization structure of the nebula precludes a simple ICF. However, if we assume that NGC 3199 has a similar total log(C/H) value to the Galactic H~{\sc ii} region Orion (8.37; \citealt{2011AA...526A..48S}), it would imply that only 0.25\% of the total gas-phase carbon was in the neutral state.

\subsection*{RCW 58 (WR 40)}

RCW 58 around the WN8 star WR 40 displays a spectrum with much in common with the other low-excitation Galactic WR nebulae in our sample. Strong [N~{\sc ii}] features are combined with weak [O~{\sc iii}] lines and a clear He~{\sc i} spectrum.

From Figure~\ref{NO_range_plot} it is clear that RCW 58 has a very high measured log(N/O), lying in the range 0.1 $<$ log(N/O) $<$ 0.55. The lower limit on this value (assuming a factor of two uncertainty) is still very high. This implies that the material possesses a CNO-cycle processed stellar ejecta component. 

This finding agrees with previous results, however the analyses of \citet{K84} and \citet{1990ASPC....7..135R} relied upon the method of tuning the electron temperature until the ``correct" oxygen abundance was achieved (in this case the Galactic ISM oxygen abundance). This methodology may be inappropriate for nebulae which are believed to contain processed material, as the true oxygen abundance is likely to differ from the ISM oxygen abundance. \citet{K84} found log(N/O) = -0.42 using an electron temperature of 7500 K while \citet{1990ASPC....7..135R} found log(N/O) = -0.3, however \citet{1990ASPC....7..135R} did not include a list of line detections, or finding charts indicating which part of RCW 58 they observed so it is impossible to draw further comparisons. In both cases the difference in derived log(N/O) stems from the difference in the strength of the [O~{\sc ii}] 3727, 3729\AA\ doublet. We found a dereddened [O~{\sc ii}] doublet intensity of $\sim$ 42 (relative to H$\beta$ = 100), while \citet{K84} quote 172, a factor of four greater. The \citet{K84} observations were not sky-subtracted, which may go some way to explain this discrepancy.

\section{Conclusions}

We found evidence of significant nitrogen abundance enhancements in the nebula around WR 8, a newly discovered nebula which we had morphologically categorised as an ejecta type nebula \citep{2010MNRAS.409.1429S}. The Galactic nebulae, with the exception of NGC 3199, were found to have significantly enhanced N/O ratios compared to M17. The nebulosity around WR 40 (RCW 58) was the most enriched (log(N/O) $>$ 0.1), followed by that around WR 16 (log(N/O) $>$ -0.1), both of which have WN8h exciting stars. The nebula around WR 8 (WN7/WC4) is also significantly enriched in nitrogen (log(N/O) $>$ -0.4). 

The N/O ratios derived for all of the LMC nebulae were consistent with LMC H~{\sc ii} region values (log(N/O) = -1.49). 

Our very high resolution UVES data allowed us to determine the NGC 3199 spectral line widths very precisely. In combination with correction factors for H$\alpha$ published by \citet{1999A&AS..135..359C}, these widths were used to measure the nebular temperature by determining the degree of thermal broadening thought to be affecting each line. The temperatures determined using this method for different ions were systematically lower than those determined using empirical line ratio temperature indicators for the same ions. Higher electron temperatures were found for low ionization ions -- an effect we attribute to radiation hardening. 

A secondary goal of this project was to determine whether carbon abundances could be derived for nebulae around carbon rich WR stars. Insufficient spectral resolution precluded us from detecting [C~{\sc i}] lines from the WC nebulae which we targeted. However we have shown that the far-red [C~{\sc i}] lines could be detected from NGC 3199 in a relatively modest amount of time with a suitably high resolution instrument. 

We did not detect any heavy element recombination lines in the UVES spectra of NGC 3199. Since these lines lie mainly in the blue this was likely due to the high reddening towards this Galactic plane target. A similar set of UVES observations of suitable nebulae in the LMC could potentially yield detections of the far-red [C~{\sc i}] lines and the C~{\sc ii} 4267\AA\ recombination line, which could allow a  quantification of the carbon abundances in WR nebulae.

The lack of appropriate temperature diagnostics for several nebulae rendered the derivation of elemental abundances for them impossible. The non-detection of the [O~{\sc iii}] 4363\AA\ line in several Galactic WR nebulae was due to several effects: low spectral resolution, high reddening and low O$^{2+}$ abundances. Observations at higher spectral resolving powers, similar to those conferred by UVES, can enable the detection of crucial temperature diagnostic lines, along with the above-mentioned diagnostic lines of carbon.

\bibliographystyle{mn2e}

\label{lastpage}

\end{document}